\documentstyle[12pt,preprint]{aastex}

\begin{document}

\title{Spectral Structure in FRII Radio Galaxies and Jets}

\author{Kari Treichel and L. Rudnick}
\email{kari@astro.umn.edu, larry@astro.umn.edu}
 
\affil{Department of Astronomy, University of Minnesota, 116 Church
Street SE, Minneapolis, MN 55455}

\author{M. J. Hardcastle}
\affil{H.H. Wills Physics Laboratory, University of Bristol, Royal
Fort, Tyndall Avenue, Bristol BS8 1TL, UK}
\email{M.Hardcastle@bristol.ac.uk}
\and
\author{J. P. Leahy}
\affil{Jodrell Bank Observatory, University of Manchester, Macclesfield, Cheshire SK11 9DL,UK}
\email{jpl@jb.man.ac.uk}
\author{REVISION:  18 May, 2001}

\begin{abstract}

Using spectral tomography to separate overlapping spectral features in 
a sample of FRII radio galaxies, we find a variety of spatial/spectral 
features that are not easily described in the context of current models.
In particular, we find mixtures of flat and steep spectrum features 
in the hot spot regions, and non-monotonic spectral index gradients 
along jets.  Additional findings include spectral gradients in 
compact hot spots and possible transitions in jet properties at the 
downstream ends of the diffuse lobes.  The complexity of behaviors 
uncovered here points to the need for a thorough investigation of 
numerical models for radio galaxies, as well as detailed 
observational studies of larger, unbiased samples.  We also perform 
the first quantitative assessment of errors in the use of spectral 
tomography for spectral index measurements.

\end{abstract}
 

\section{Introduction}

Radio source spectral indices are used to probe the gain and loss
mechanisms of the relativistic electron population in extragalactic
radio sources.  Steeper spectra may indicate that the electrons in the
source have ``aged'' through synchrotron and other losses.  Flatter
spectra can be caused by {\it in situ} particle acceleration or
compression of the medium. (Leahy, 1991 and Eilek \& Hughes, 1991).
The spectral index variations in a source are therefore fundamental
for testing models for the evolution of the relativistic particle
populations in radio galaxies.  The relative importance of acceleration
at the jet terminus, or along the jet itself, or in the more diffuse
lobes is indeterminate at present.  Challenges to the standard picture
from recent numerical simulations (e.g., Tregillis, Jones \& Ryu, 2001)
 and from 
non-thermal emission in optical and X-ray bands (e.g., Celotti, Ghisellini \& Chiaberge, 2001,
Wilson, Young \& Shopbell 2001) raise questions about the origins of
spectral variations in the much lower energy radio spectra.

Unfortunately, the observation and interpretation of spectral variations
in radio galaxies is often ambiguous (e.g., Rudnick, 2001a). 
 One problem is that the observed  
spectral index of a feature in a radio source is  misleading  when
two structures with different spectral indices overlap in the plane of
the sky.  For example, flatter spectrum jets often overlap steeper
spectrum lobes.  Spectral tomography is one way around this problem
where  a feature with a specified spectral index can
be subtracted out of the image leaving only the second structure.

Spectral tomography has been used to analyze the spectra of several
Wide Angle Tail sources (Katz-Stone \& Rudnick, 1997b and Rudnick \&
Katz-Stone, 1995) and Compact Steep Spectrum sources (Katz-Stone \&
Rudnick, 1997a). Tomography of one FRII, Cygnus~A, isolated features
similar to those seen in the ``frequency-corrected" images of this
prototypical source (Rudnick, Katz-Stone \& Anderson, 1994).
In each case, features that were unexpected from current radio galaxy
models were found.  We therefore initiated this exploratory study of
additional FRII radio galaxies using existing maps.  Below we examine
the results from our spectral tomography on these distant, FRII
sources and estimate the spectral index of several jets along their
lengths.  We then briefly describe some implications of these results
for radio galaxy models, and make suggestions for future work.

\section{Observations}

\begin{figure}  
\centerline{
\plotone{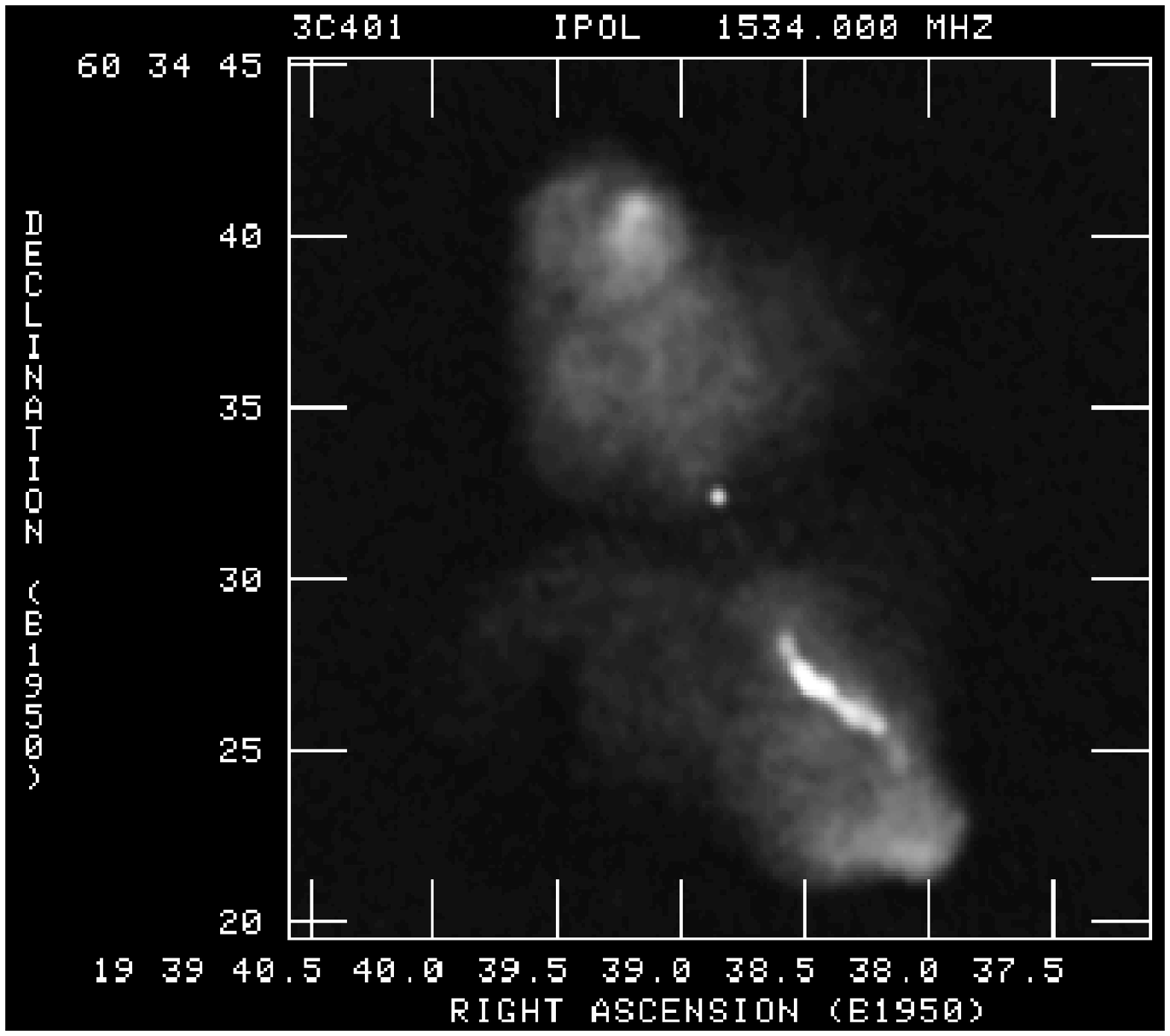}
}
\caption{Total intensity map of 3C401 at 1.5 GHz.}
\label{grey401}
\end{figure}

\begin{figure} 
\centerline{
\plotone{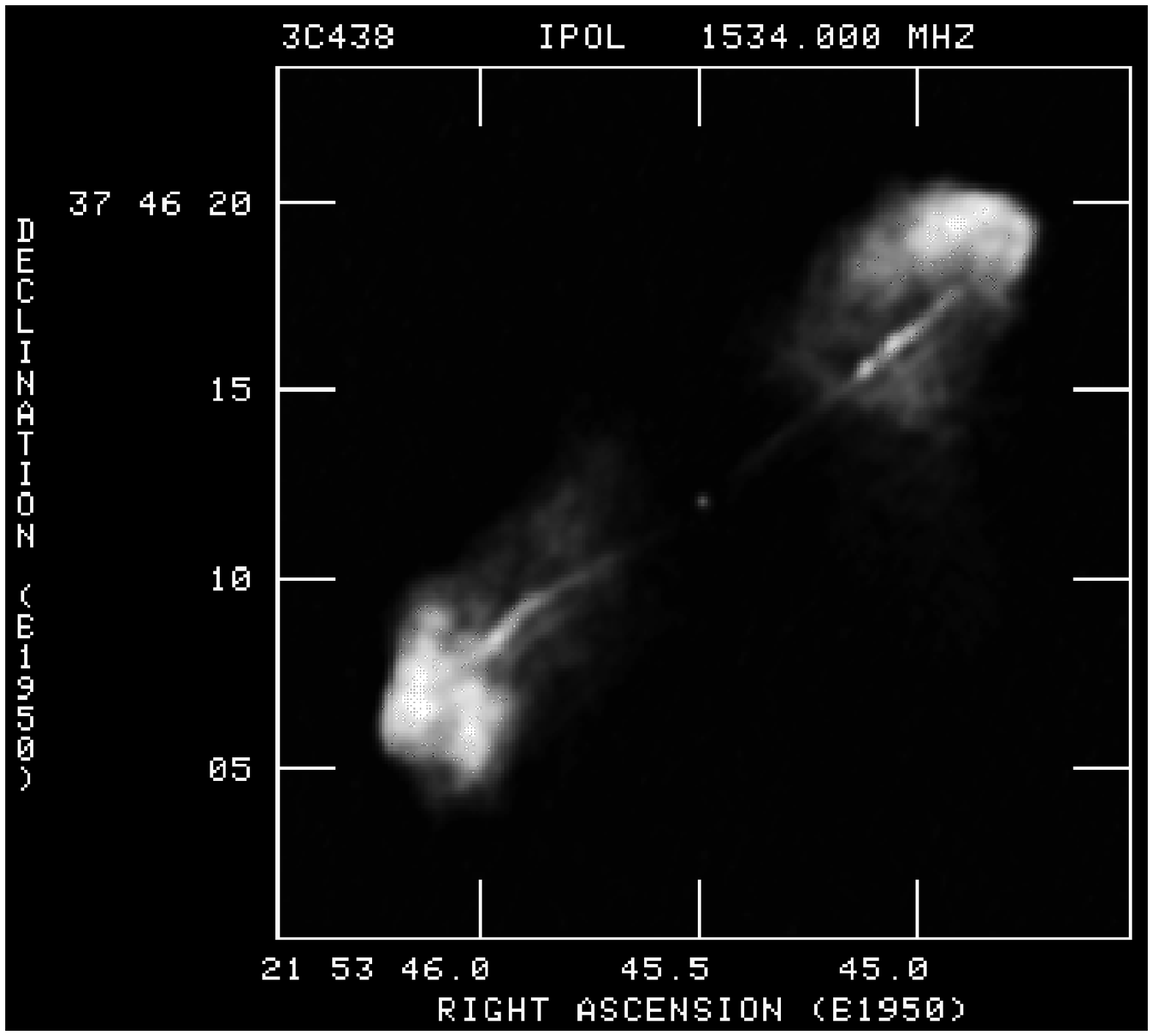}
\vspace{12pt}
}
\caption{Total intensity map of 3C438 at 1.5 GHz.} 
\caption{Spectral tomography maps of 3C438 showing the mixture of steep
and flat spectral indices in each lobe. The units of these maps are
Jy/beam, as were the original intensity maps.  The off-source grey regions
have means near zero and fluctuations that give an indication of the
noise.  Bright white regions are those that are flatter than $\alpha=
0.81,1.02$ for the top, bottom, respectively. Dark blacker regions
are steeper than these respective values.  Fine scale mottling in the
steep regions is an artifact of the X-band map reconstruction.  }
\label{grey438} 
\end{figure}

The maps used in this study were all obtained from previous studies.
The maps at 8.4 GHz are VLA \footnote{The Very Large Array is a
facility of the National Radio Astronomy Observatory, operated by
Associated Universities, Inc., under contract with the National
Science Foundation.} data from Hardcastle et al. (1997).  The maps of
the sources at a second frequency come from Leahy and Perley (1991)
for 3C173.1, 3C349, 3C381, and 4C14.11 (VLA data); from Leahy et
al. (1997) \footnote{Available at URL: http://www.jb.man.ac.uk/atlas}
for 3C153, 3C401, and 3C438 (VLA and MERLIN data) \footnote{MERLIN is
a national facility operated by the University of Manchester on behalf
of the UK Particle Physics and Astronomy Research Council (PPARC)};
and from Laing (private communication) for 3C132 (VLA data).  Table 1
contains a list of each source at both frequencies used and the
beamsize and noise of each map.  For all maps, the center of the uv
plane was well sampled, so the flux in the extended structure was well
constrained.  Total intensity maps of the two sources with prominent
jets, 3C401 and 3C438, are shown in Figures \ref{grey401} and
\ref{grey438}. Tests of map fidelity and the robustness of the tomography
results are discussed in the following section.

\section{Analysis and Results}
\subsection{Map preparation}

Prior to constructing the spectral tomography galleries, we first
convolved the maps at each frequency to the same resolution.  Then the
maps were aligned by matching up the peak of the core or a hot spot (if
the core was absent from the images).  The peak locations were found by
fitting the core or hot spot to elliptical gaussians.  These alignments
were verified by subtracting the maps at the two frequencies with small
relative shifts in position and minimizing the rms residuals in a
small area surrounding and including the core or hot spot.  Any
misalignment would result in a gradient in spectral index seen
throughout the spectral tomography maps.  No such global gradient was
seen, and the alignments are correct to within 0.1 pixels.

\subsection{Map fidelity}

Because our work was based on previously published maps, constructing the
spectral tomography galleries involved comparing images which had normally
been imaged and deconvolved using different algorithms or weightings. In
addition, some of our pairs of maps were made from observations which had
very different samplings of the $uv$ plane. It was important to be certain that
these differences did not affect the reliability of the tomography images for
the structures of scientific interest.

To assess the effect of these differences, we returned to the original $uv$
data for the two jetted sources 3C\,401 and 3C\,438. The 8.4-GHz VLA data for
these sources are a combination of observations with the A, B and C
configurations of the VLA, with no more than an hour's observation at each
configuration (together with a 10-minute snapshot at D-array in the case of
3C\,438). At our second frequency, 1.4 GHz, the data consist of a long track
with MERLIN together with short observations with the A configuration of the
VLA. While the longest and shortest baselines sampled by these observations
are very similar, the detailed coverage of the $uv$ plane is quite
different. Because of the uneven sampling of the $uv$ plane, the MERLIN+VLA
data were particularly prone to the `rice pudding' effect introduced by
instabilities in the CLEAN algorithm (Cornwell 1983), and so they were deconvolved
using a hybrid CLEAN/MEM technique (e.g. Leahy \& Perley 1991), whereas the
8.4-GHz VLA data were simply deeply CLEANed. Because of these differences, and
because of the complex structure seen in the two sources, they are good
test cases for our investigation.

We quantified the effects of different sampling and deconvolution by
simulating observations of the {\it same} source using the two different
datasets. To do this we took the CLEAN components from a deep CLEAN of the
8.4-GHz data for the two sources at full resolution, assuming that this map
represents adequately the total-intensity structures of interest. Using the
{\sc aips} task UVSUB, we replaced the real visibilities in both 1.4 and
8.4-GHz $uv$ datasets with a model based on the CLEAN components. We then
mapped the datasets using deconvolution techniques that were similar to those
used to produce the real maps, convolved to a common resolution, and compared
the resulting images.

If the deconvolution had been ideal, and the sampling perfect, we would expect
to obtain identical maps from this process. In fact, there were small
differences, particularly in the low-surface-brightness structure, which
presumably result either from different sampling or, more likely, from
artifacts introduced by deconvolution (such as rice-pudding effects in the
deeply cleaned simulated 8.4-GHz maps). However, there were no {\it systematic}
differences in the sources, which would reflect serious undersampling, and the
high-surface-brightness regions of the simulated sources were essentially
identical in the two maps. When we constructed maps of `spectral index'
between the two maps, the r.m.s. variation was of order 0.05 in the
high-surface brightness regions, and the mean was indistinguishable from
zero. We are therefore able to conclude that the differences in $uv$ plane
sampling and deconvolution do not introduce any significant errors into the
analysis that follows; by contrast, the {\it detailed} structure of
low surface brightness tomography images is not reliable and not
discussed here.

The deconvolution procedures recovered most of the expected total 
fluxes.  The maps contain approximately 90\%, 100\%  at 1.4, 8.4 GHz, 
respectively, for 3C401, and 95\%, 100\% for 3C438.  Any  missing 
fluxes, which are likely to correspond to the largest-scale 
components, would not significantly affect any of the results in this
paper.  Tomography of well-sampled small-scale features, as discussed below, 
effectively removes any biases due to underlying diffuse structures, 
so errors in the reconstruction of the latter do not propagate.  This 
would not be true, by contrast, with the standard method of 
calculating spectral indices using map division. 

\subsection{Spectral Tomography}

To take a closer look at overlapping structures, one wants to first subtract
out of the image a feature of a particular spectral index. Since we do
not know {\it a priori} which spectral index to use to remove confusing
structures, we construct a gallery of maps, each with a different
spectral index ``zeroed'' out.  That is, we construct a set of maps
$M_{\alpha_{t}}$ where,

\begin{center}  $M_{\alpha_{t}}=M_{\nu_1}-M_{\nu_2} * \exp(\alpha_t \ln {\frac{\nu_1}{\nu_2}})$. 
\end{center}

\noindent If a structure has spectral index $\alpha_{t}$, the structure will
disappear in map $M_{\alpha_{t}}$. 

The intensity of the signal that is left behind in a tomography image
has two origins - its overall intensity and a factor which depends on
the difference between its actual spectral index and $\alpha_{t}$ .  We quantify this as follows: Suppose a feature has
intensities $A_{\nu_{1}}$ and $A_{\nu_{2}}$ and a spectral index
$\alpha_{0}$ that is different than that of a particular tomography
image $M_{\alpha_{t}}$.  Then its intensity $A_{t}$ in that tomography
image will become:

\begin{center} $A_{t} = 
A_{\nu_{2}}-({\frac{\nu_{2}}{\nu_{1}}})^{\alpha_{t} }A_{\nu_{1} } = 
S_{\nu_{1} }*( \frac{\nu_{2}}{\nu_{1}})^{\alpha_{0} }[(\frac{\nu_{2}}{\nu_{1}})^{(\alpha_{t}-\alpha_{0}) } -1]$
\end{center}

Thus, the presence of a feature in a tomography image shows that it 
has a different spectral index than the corresponding $\alpha_{t}$. 
However, its strength is the product of both its original intensity and a
function depending on the difference between $\alpha_{t}$ and $\alpha_{0}$.
One cannot deduce the spectral index of a feature from tomography 
images, {\it except where it disappears}.

Because spectral tomography maps are a linear
combination of total intensity maps, they have the same reliability
and noise characteristics of the original intensity maps.  Therefore,
in the absence of deconvolution or other systematic problems, the
off-source noise can be used as an indicator of the reliability of
individual features on the tomography map.  Deconvolution problems,
such as described above, can be present in the same way on intensity
or tomography maps.  However, such problems are often more visible on
tomography images; this is because a large fraction of the signal has
been subtracted, while the artifacts at the two frequencies are
unlikely to be identical and therefore do not subtract cleanly.  We
will refer back to these guidelines in our discussion of findings from
the tomography images.

\subsubsection{Mixed Spectral Indices in 3C401 and 3C438}

\begin{figure}  
\centerline{
\plotone{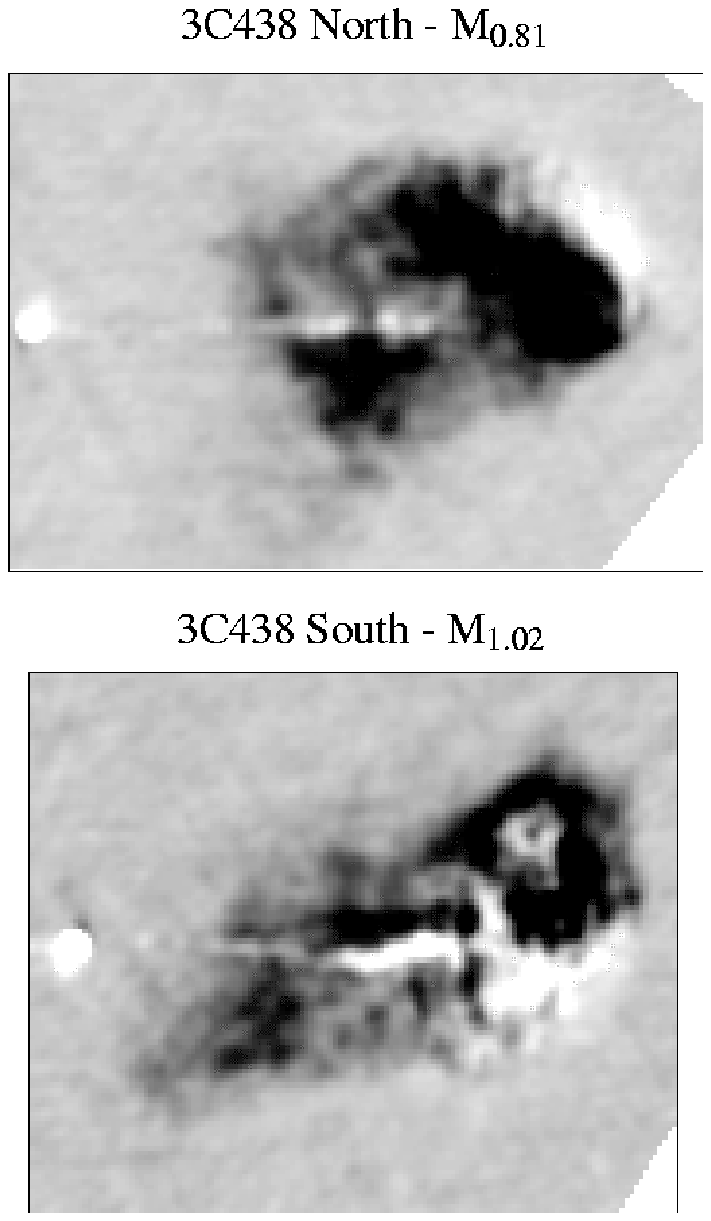}
}
\label{tomo438} 
\end{figure}

When looking through the galleries, the first thing one notices is the
mixture of spectral indices throughout the lobes of the sources.  The
most dramatic example is in source 3C438 as shown in Figure
\ref{tomo438}.  Figure \ref{tomo438} shows two spectral tomography
maps, the north lobe of 3C438 at $\alpha_t=0.81$, and the south lobe at
$\alpha_t=1.02$.  In these maps, lighter tones indicate regions with
spectral index flatter than the respective above values of $\alpha_t$
while dark regions are steeper than $\alpha_t$.  For example, the jet is
flatter than $\alpha_t$ in both these maps and shows up white.  In the
southern lobe, one can see a dark ring that has a steeper spectral
index.  There is also a light, therefore flat, ``c''-shaped structure
to the west.

\begin{figure}  
\centerline{
\plotone{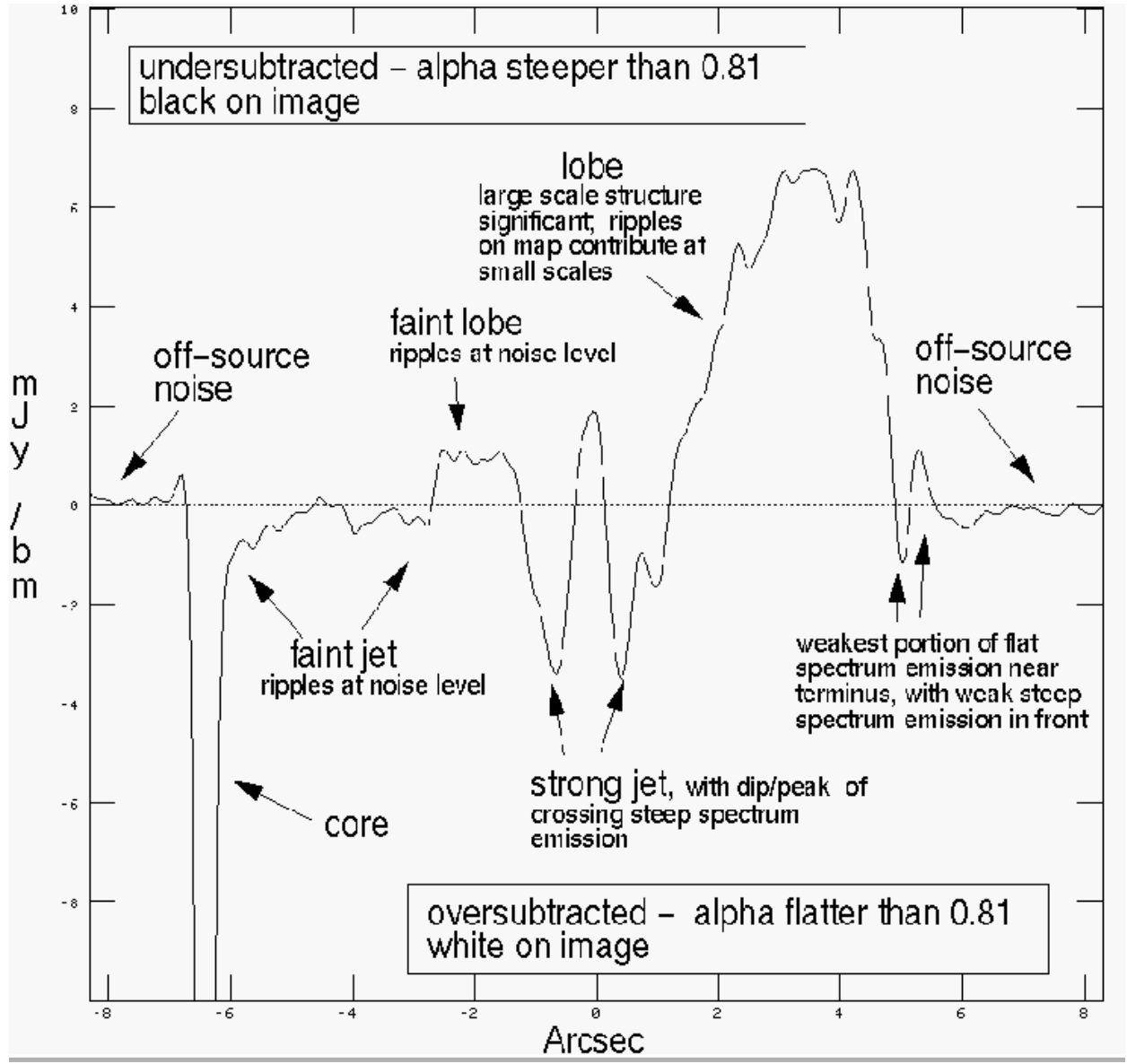}
}
\vspace{12pt}
\caption{A slice along the northern jet of 3C401 from the $\alpha=0.81$
previously shown tomography image, commenting on the reliability of
various features.   }
\label{slicetomo438} 
\end{figure}

Regarding the reliability of features, the off-source ripple indicates
the level of background noise.  Additional spurious details are seen
in the lobe emission from likely CLEAN artifacts, which can be
recognized and assessed as they would in any other total intensity
image.  The existence of the separate flat/white and steep/black
features are clear, high signal-to-noise results.  The detailed
small-scale fluctuations in the lobes are not.  A slice along
the N jet from the Figure \ref{tomo438} tomography image is shown
in Figure \ref{slicetomo438} and details the noise trustworthiness
of the various features.

3C401 and 3C153 had complex spectral structures similar to those in 3C438.  The other
sources had spectral gradients in their hot spots (see below).
but lacked sufficient resolution or dynamic range to show more than an
overall steepening of the lobe going from the hot spots towards the core.

\begin{figure}
\centerline{
\plotone{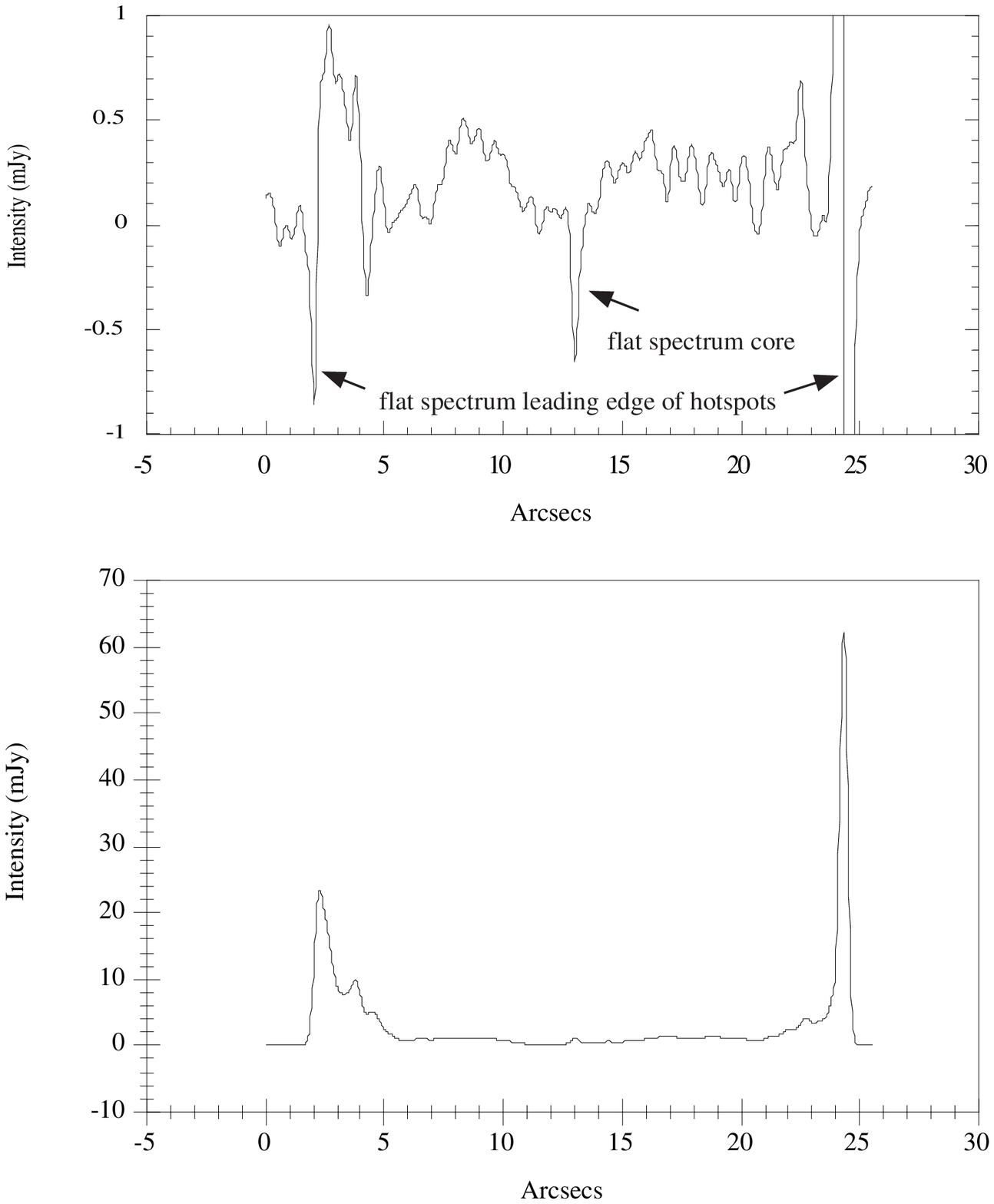}
}
\caption{a) Spectral tomography slice and b) total intensity of a slice
along the length of 3C132 from hot spot to hot spot.}
\label{slice132}
\end{figure}

When examining the spectral tomography maps, we noticed that the hot
spots appeared to have a spectral index gradient with $\alpha$ of the
outer edges of the hot spots being about 0.2 flatter than that of the
inner edges.  The first assumption upon seeing this was that the maps
were misaligned, since a spectral index gradient can be a product of
misalignment.  But if the maps were misaligned, the gradient would be
in the same direction in both hot spots, not in opposite directions
as we see in Figure \ref{slice132}.  In this figure of a slice across a
tomography map, $M_{0.78}$, intensities less than zero indicates
regions with $\alpha < 0.78$ and intensities greater than zero
indicate $\alpha > 0.78$.  Out of the eight sources we studied, four
sources had no well defined hot spots or confusing structures near the
hot spots.  In the remaining four sources with clear hot spots, we saw
a spectral index gradient clearly in three sources (3C132, 3C349, and
3C381) and tentatively in the fourth (3C173).

\subsubsection{Structure Parallel to Jet}

\begin{figure}
\centerline{
\plotone{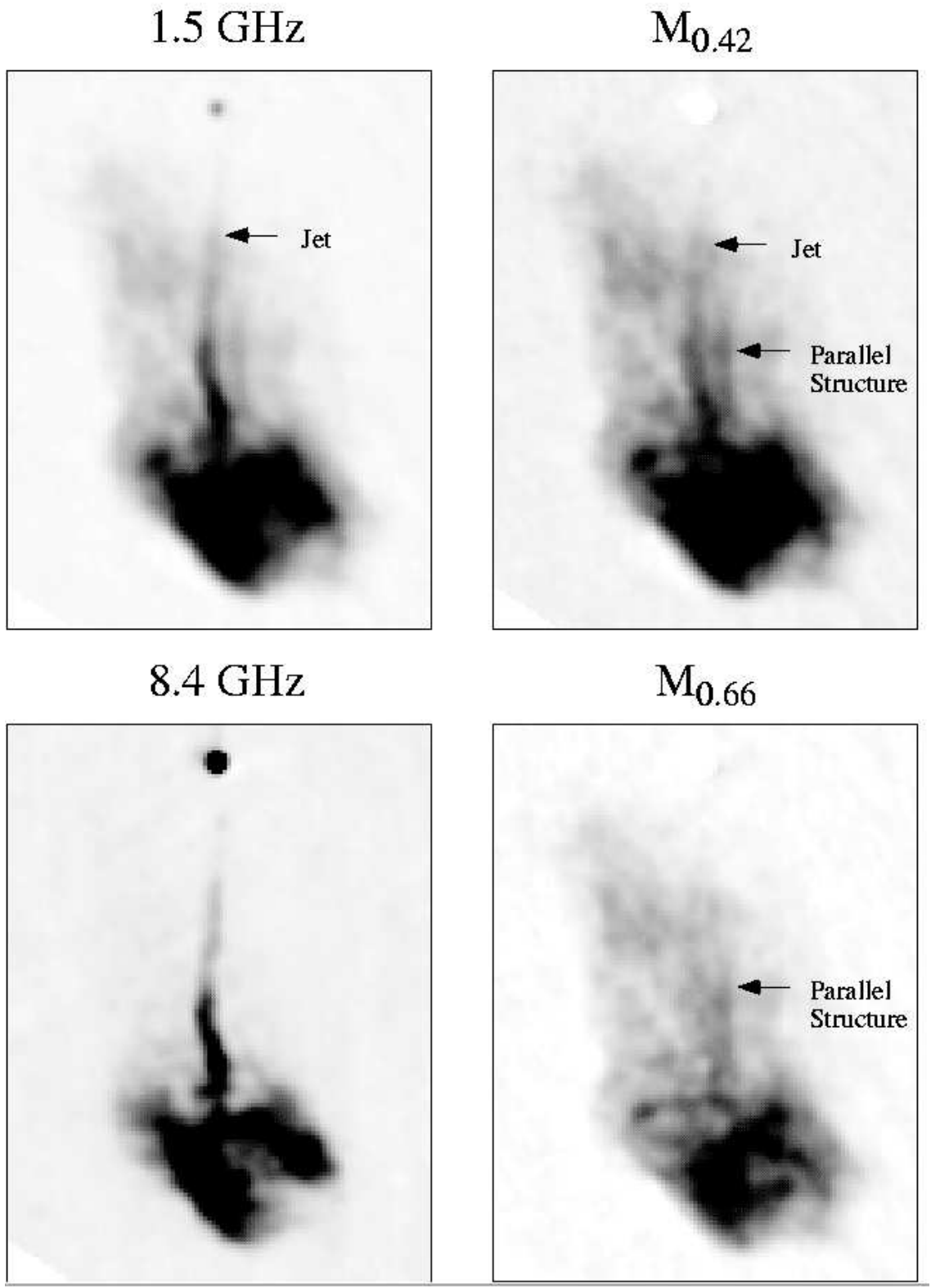}
}
\vskip -0.5in
\vspace{12pt}
\caption{Grey scale total intensity images and spectral tomography maps of 
the south lobe of 3C438 showing the structure running parallel to the jet.}
\label{parallel438}
\end{figure}

Among these structures of differing spectral index, one can see a
structure that runs parallel to the jet in 3C438.  This structure has a
steeper spectral index ($\alpha=1.2$) than the jet, as is shown in
Figure \ref{parallel438}.

\begin{figure}
\vskip -3in
\centerline{
\plotone{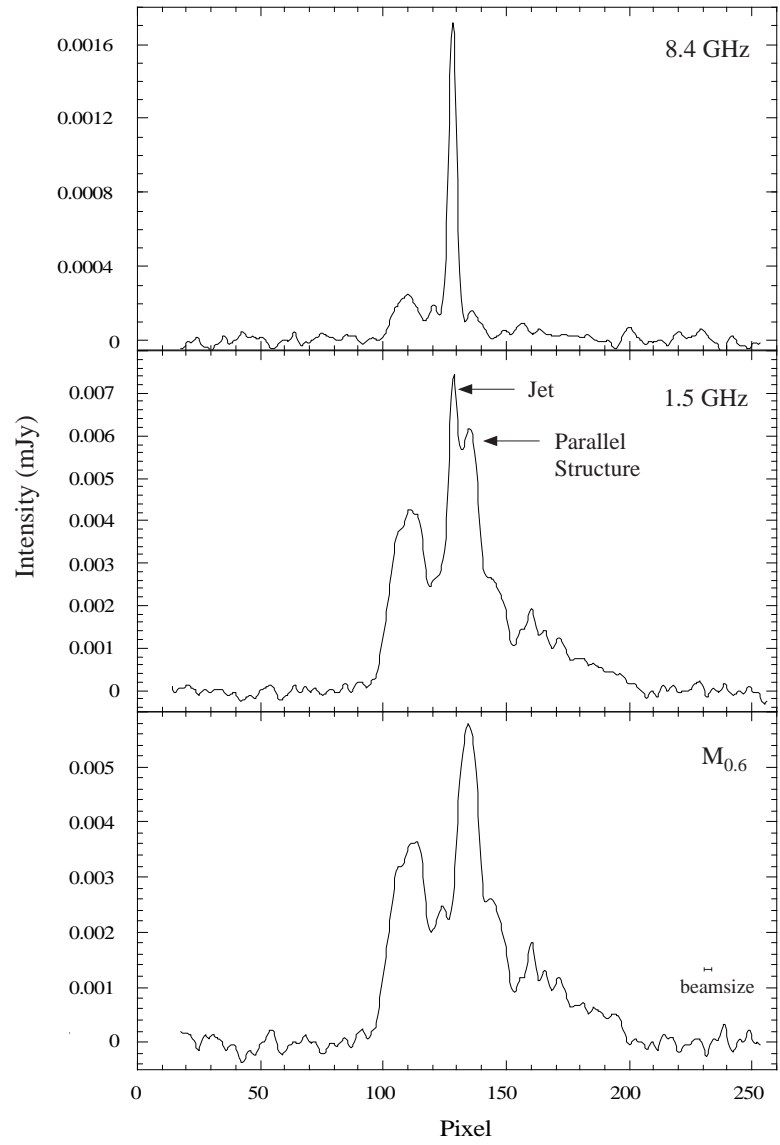}
\vspace{12pt}
}
\caption{Integrated slices across the northern jet of 3C438 showing the
parallel structure.}
\label{sliceN438}
\end{figure}

To take a closer look at this structure, we compressed the
approximately 4" in the northern lobe that included the parallel
structure to create an integrated slice across the jet.  This
integrated slice is shown in Figure \ref{sliceN438} for 8.4 GHz,
1.5 GHz, and for $M_{0.6}$, where the jet has disappeared from the
tomography maps.  The parallel structure is easily seen in the 1.5 GHz
map, but is weaker in the 8.4 GHz map.  The slices show that there may
also be a steep spectrum feature partially blended with the jet on the
opposite side from the parallel structure.

\begin{figure}
\centerline{
\plotone{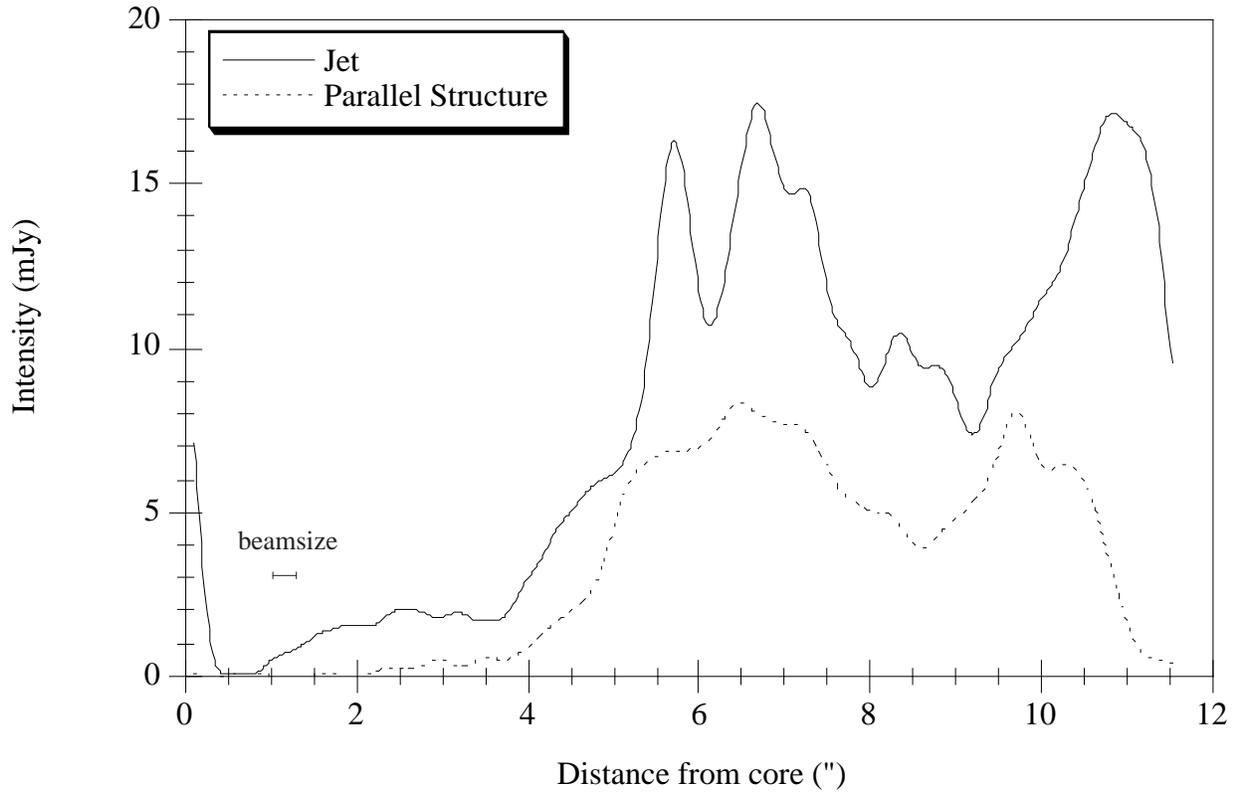}
}
\vspace{12pt}
\caption{Slices along the jet and parallel structure in 3C438 at 1.5
GHz showing similar but not identical behavior.  The rms noise for
this image is 97 $\mu$Jy.}
\label{slicepar438}
\end{figure}

Figure \ref{sliceN438} clearly shows that the structure we 
examined consists of a single peak, not multiple peaks or a sinusoidal pattern
which would be expected if the structure was simply an imaging
artifact.  To further ensure that the structure was not an imaging
artifact, a slice was taken along the jet and along the parallel
structure.  These slices are shown in Figure \ref{slicepar438}.  The
intensity of the parallel structure does not mimic the major small-scale peaks in the jet
as would be expected for an imaging artifact.  Furthermore, a similar
structure parallel to the jet can be seen in the southern lobe where
the jet bends.  In the southern lobe, the parallel structure closely,
but not perfectly, follows that bend of the jet.  This behavior is not
likely to result from an imaging artifact.

\subsection{Error Analysis of Jet Spectra}

To obtain more quantitative results on jet spectra in the two strongly
jetted sources, 3C401 and 3C438, slices were taken across the jet for
each spectral index  $\alpha_t$ in the gallery.  When
$\alpha_{t}=\alpha_{jet}$,  the jet will disappear from the
slice.  Note, however, that the intensity at the position of the jet
in $M_{\alpha_{t}}$ does not go to zero, but instead reflects the residual
lobe intensity at $\alpha_t$.

\begin{figure}
\centerline{
\plottwo{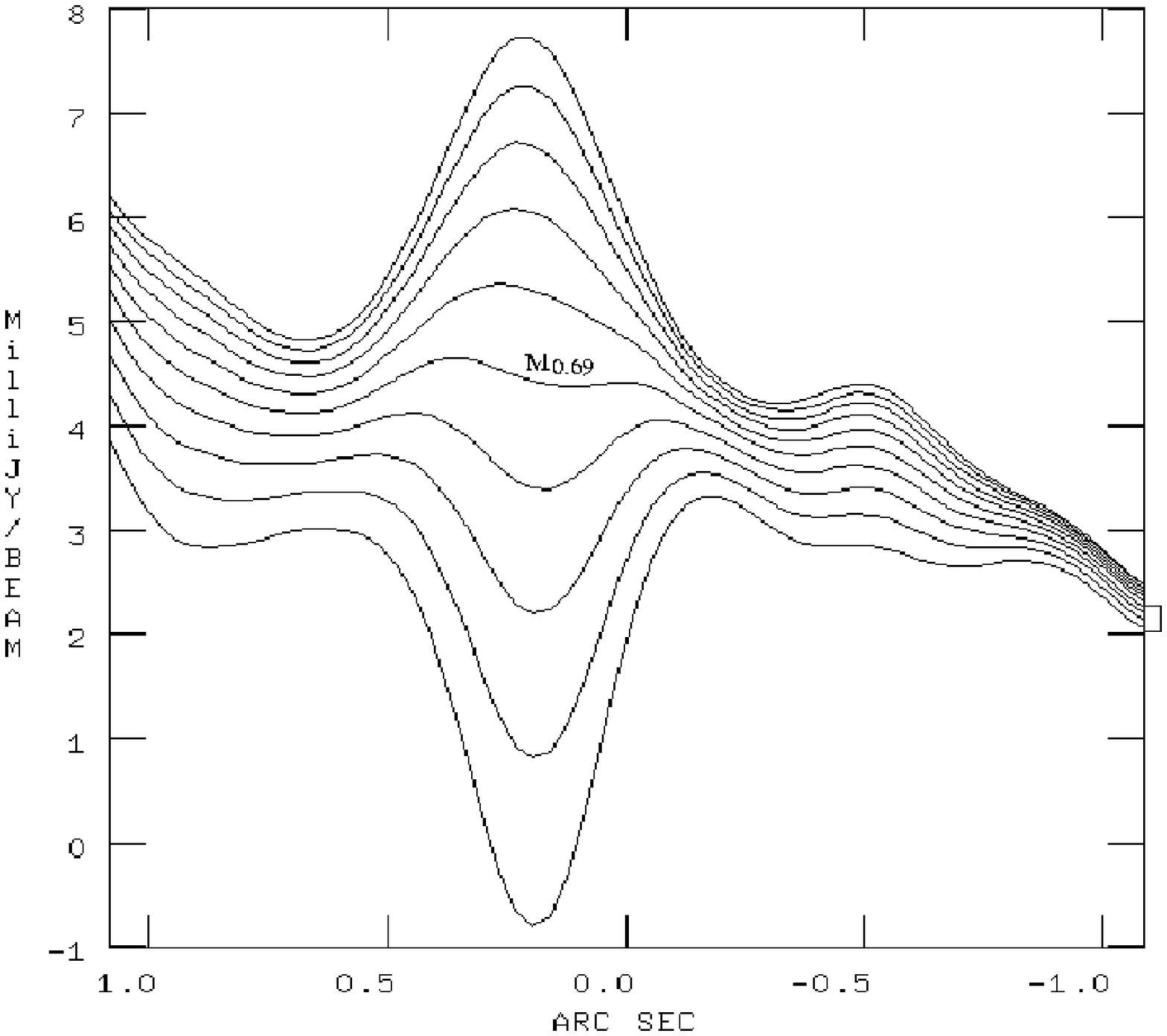}{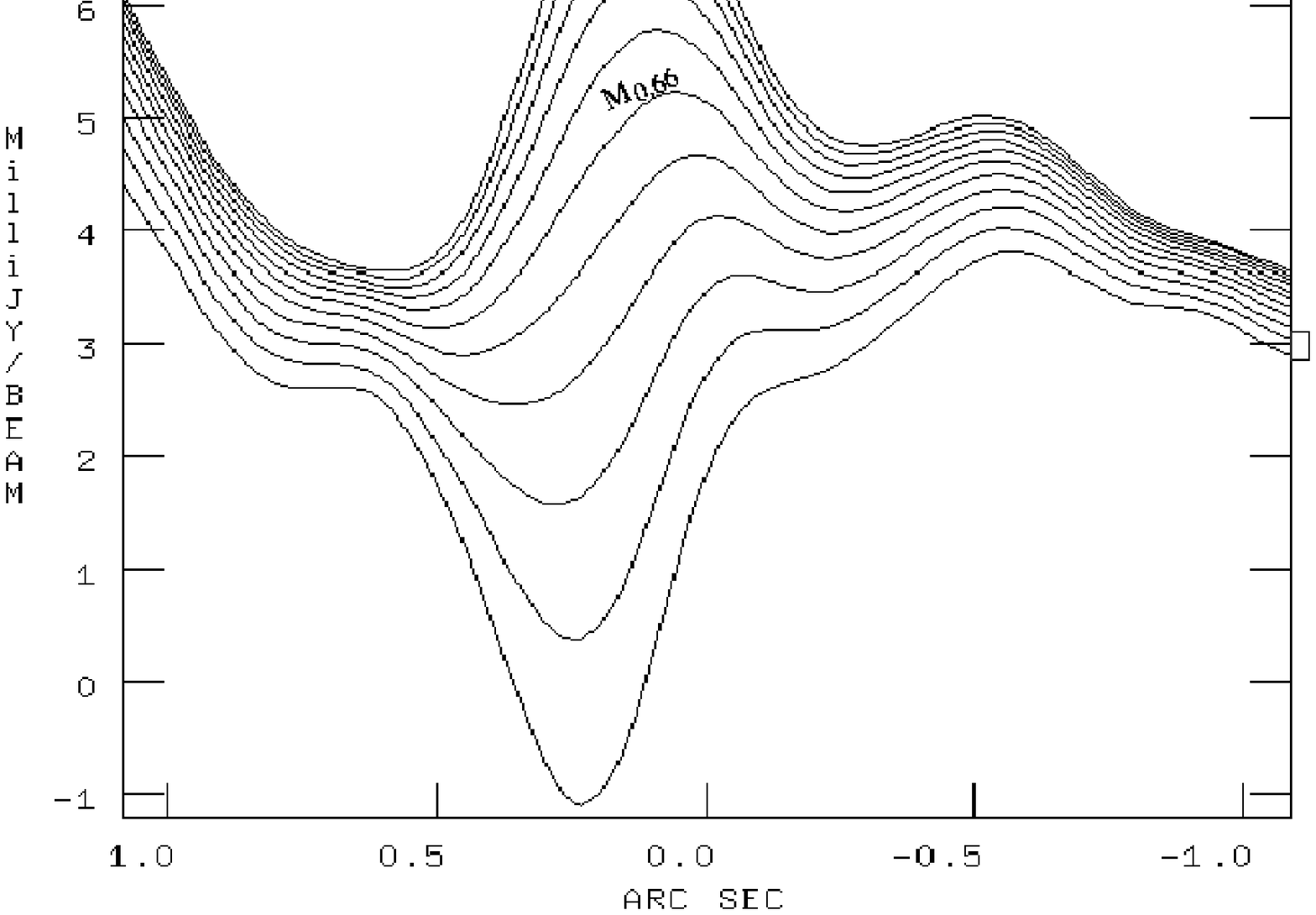}
}
\vspace{12pt}
\caption{a) A series of slices through the jet spectral tomography images for 
a range of spectral indices.  The labeled slice indicates the one identified 
by eye as corresponding to the jet spectral index.  b) Same as a) but showing
greater confusion from the noise and extended structure.}
\label{egslices}
\end{figure}

Determining $\alpha_{jet}$, when the jet disappears,  is necessarily
subjective.  In some cases, the determination is straightforward, in
others, confusion from the background (lobe) emission makes
identification of the local jet spectral index difficult.  Examples of
each of these cases are shown in Figure \ref{egslices}.  Below,
 we estimate the errors in our results caused by confusing
structures.

\begin{figure}
\vskip -3in
\centerline{
\plotone{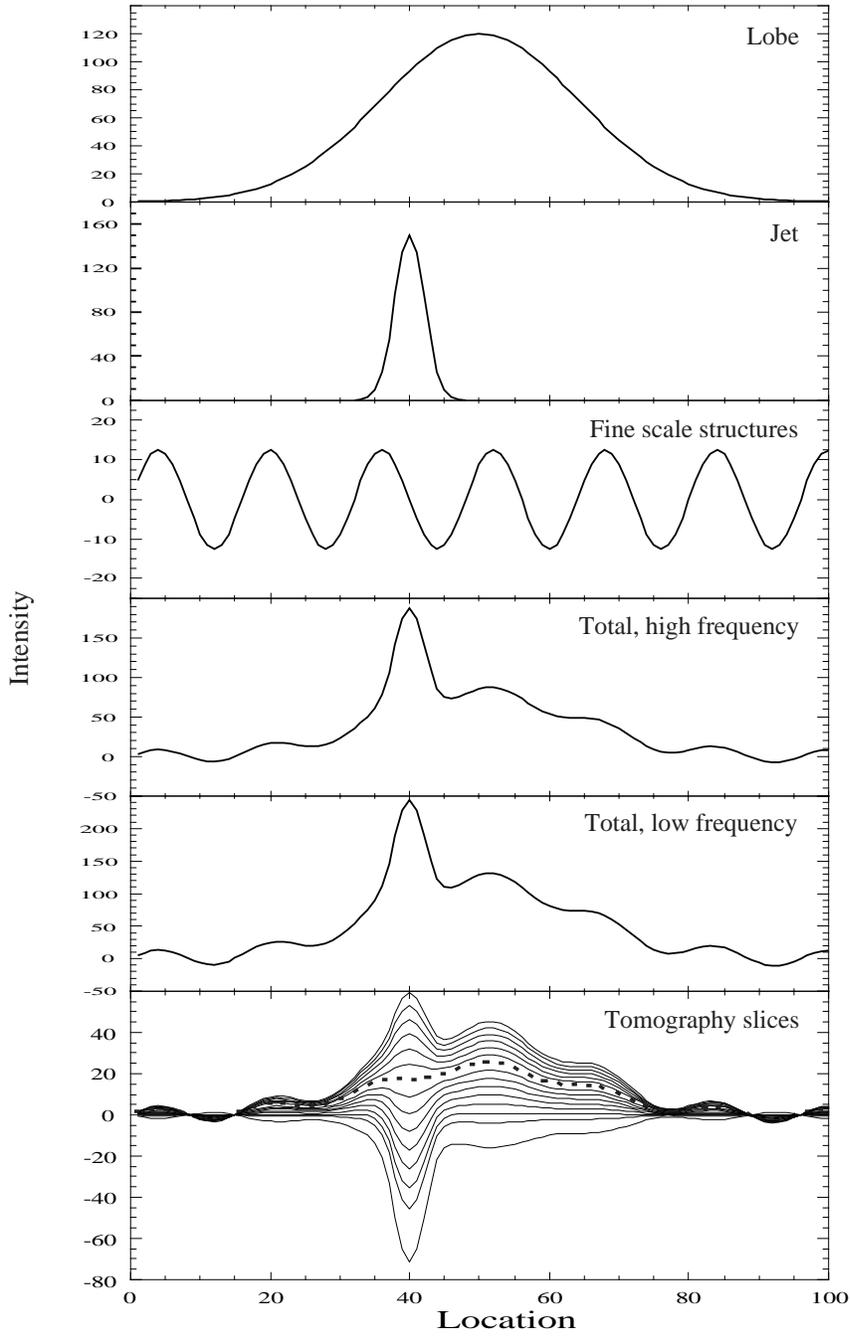}
}
\vspace{12pt}
\caption{Example of the models used to estimate the errors in spectral 
tomography determinations of spectral index.  The top three panels show the
individual components.  The next two panels show the sum of the components,
weighted by spectral index at both high and low frequencies.  The bottom 
panel shows slices through the tomography maps of different spectral indices.  
The thick dashed line is the slice corresponding to the correct spectral 
index and is also the slice we picked as the correct one.}
\label{models}
\end{figure}

To test the reliability of determining spectral indices using
tomography in the presence of confusing structures, we made a series
of simple, one-dimensional, numerical models at two frequencies with a
ratio of 1.44:1.  These models included a narrow gaussian jet (FWHM =
3), a wide gaussian lobe (FWHM = 20), and superposed sine waves of
different wavelengths to simulate irregularities in the lobe emission.
The spectral index of the jet was set to 0.5, while the lobe and
irregularities were set to 1.1 (see Figure \ref{models}).
Then we varied several characteristics of the jet and lobe including
the position of the jet relative to the lobe, and the width,
intensity, and phase of the sine wave.  Obviously, using a sine
wave to simulate random irregularities is artificial, as the eye can
detect the sine wave pattern in the unconfused area and follow it into
the confused area, i.e. the jet.  Whereas, when looking at the real
data, it is impossible to tell when a confusing structure is perfectly
overlapping the jet.  But using the sine wave gives a useful approximation
to the errors in identifying the ``correct'' jet spectral
index.

\begin{figure} 
\centerline{
\plotone{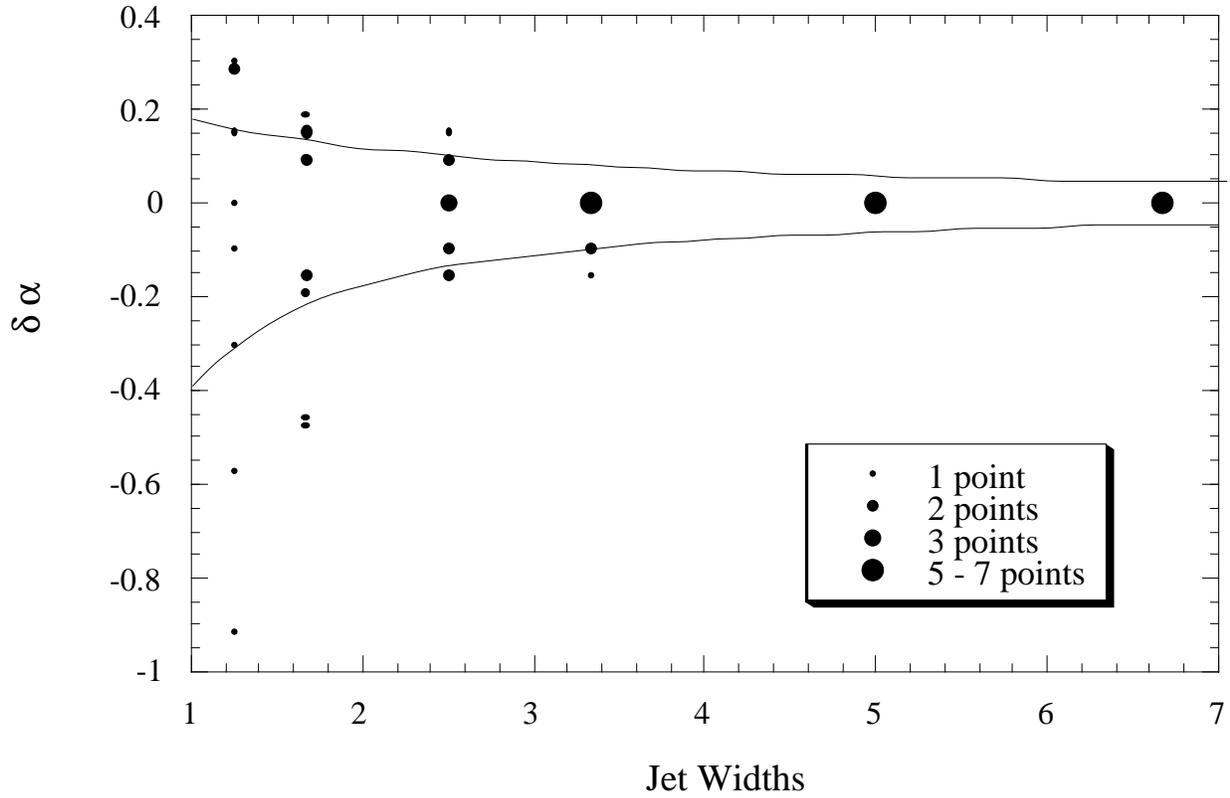}
}
\vspace{12pt}
\caption{Error in $\alpha$ from tomography of model jets vs. wavelength of 
the sine wave used to simulate noise.  The closer the width of the sine 
wave is to the width of the jet, the larger the errors in $\alpha$.  The 
solid lines are the analytical errors found using the equation in text, 
where $b_1$ was taken as a sine wave.  The size of the symbols indicates 
the number of times the error in spectral index took on that value.} 
\label{errors} 
\end{figure}

The most important effects on whether or not the correct spectral index
was determined were the interfering sine wave's width and intensity at the jet
position (i.e., the phase of the sine wave); the jet's position with respect to the lobe was not
important.  When looking at the slices, any sine waves significantly
wider than the jet were easily distinguishable; determining the
spectral index of the jet was difficult only when the width of the sine
wave was on the same order as the jet.  Figure \ref{errors} shows
how the width of the sine wave affects the size of the error in
$\alpha$. Plotted in this figure are the differences between the
correct $\alpha$ and the $\alpha$ determined by spectral tomography for
various sine waves of different wavelengths and different offsets of
the peak from the jet.  This figure shows that the correct spectral 
index falls within the {\it range mean $\pm$ error} about 2/3 of the time,
with errors calculated analytically as discussed below.
 This implies that our  calculated  errors correspond to  1 $\sigma$.

\begin{figure} 
\centerline{
\plotone{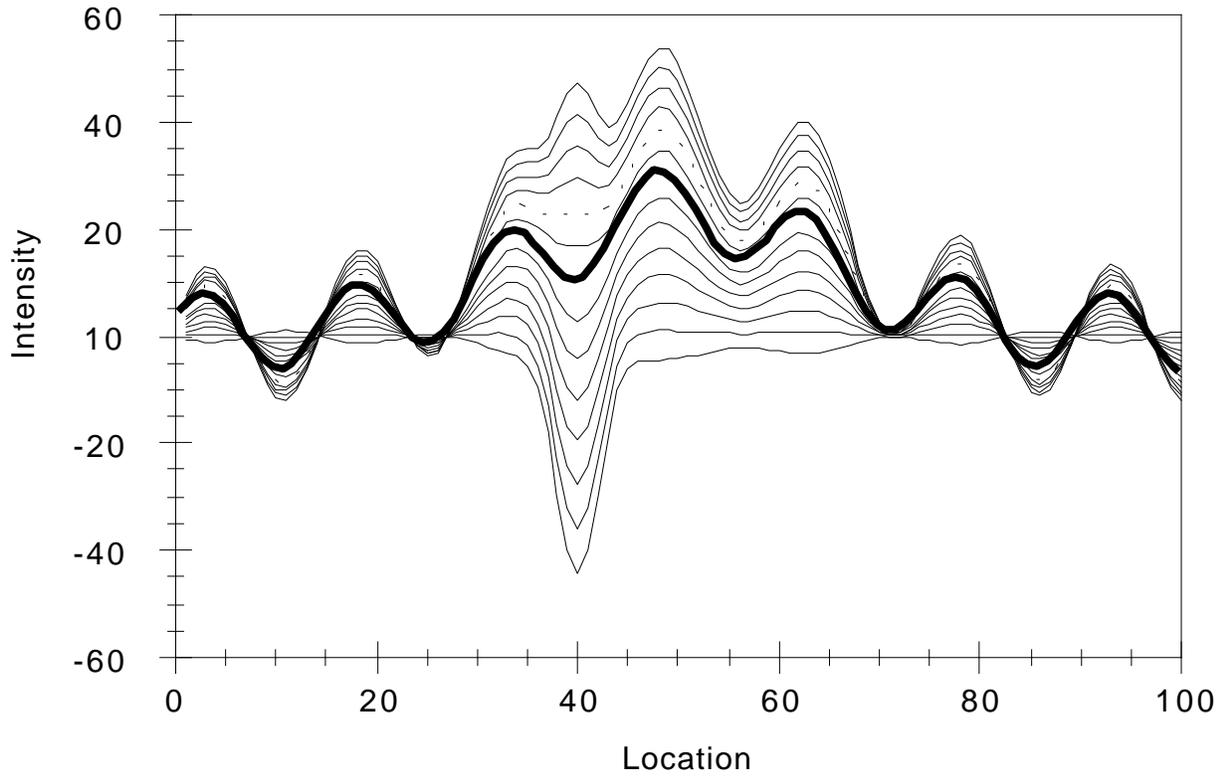}
}
\vspace{12pt}
\caption{Spectral tomography slices of a model with significant confusion 
from the sine wave ``noise''.  The thick line is the slice corresponding to 
the correct spectral index ($\alpha$ = 0.5), and the dashed line is 
the slice identified by tomography ($\alpha$ = 0.71). }
\label{noise}
\end{figure}

Figure \ref{noise} shows an extreme case where the minimum of the
sine wave occurs at the jet.  In this situation, inspection by eye gave
a spectral index which was incorrect by 0.21.  Analytically,
the same error can be found by defining the true  spectral
index as  $\alpha_{0} = {\frac {\ln(I_1/I_2)}{\ln(\nu_1/\nu_2)}}$, where $I_1$
and $I_2$ are the intensities at frequencies $\nu_1$ and $\nu_2$
respectively.  With a confusing structure of intensity $b_1$ and $b_2$,
the observed spectral index will be $\alpha_{obs} = {\frac {\ln
((I_1+b_1)/(I_2+b_2))}{\ln(\nu_1/\nu_2)}}$.  This leads to an error in
spectral index of the jet of

\begin{center}
    $\delta \alpha = \alpha_{obs}-\alpha_{0} $

$= {\frac {\ln ((I_1+b_1)/(I_2+b_2))-\ln(I_1/I_2)}{\ln(\nu_1/\nu_2)}}$

$= (\ln({\frac{I_1+b_1}{I_1}})-\ln({\frac{I_2+b_2}{I_2}}))/\ln({\frac{\nu_1}{\nu_2}})$
\end{center}

\begin{figure}
\centerline{
\plotone{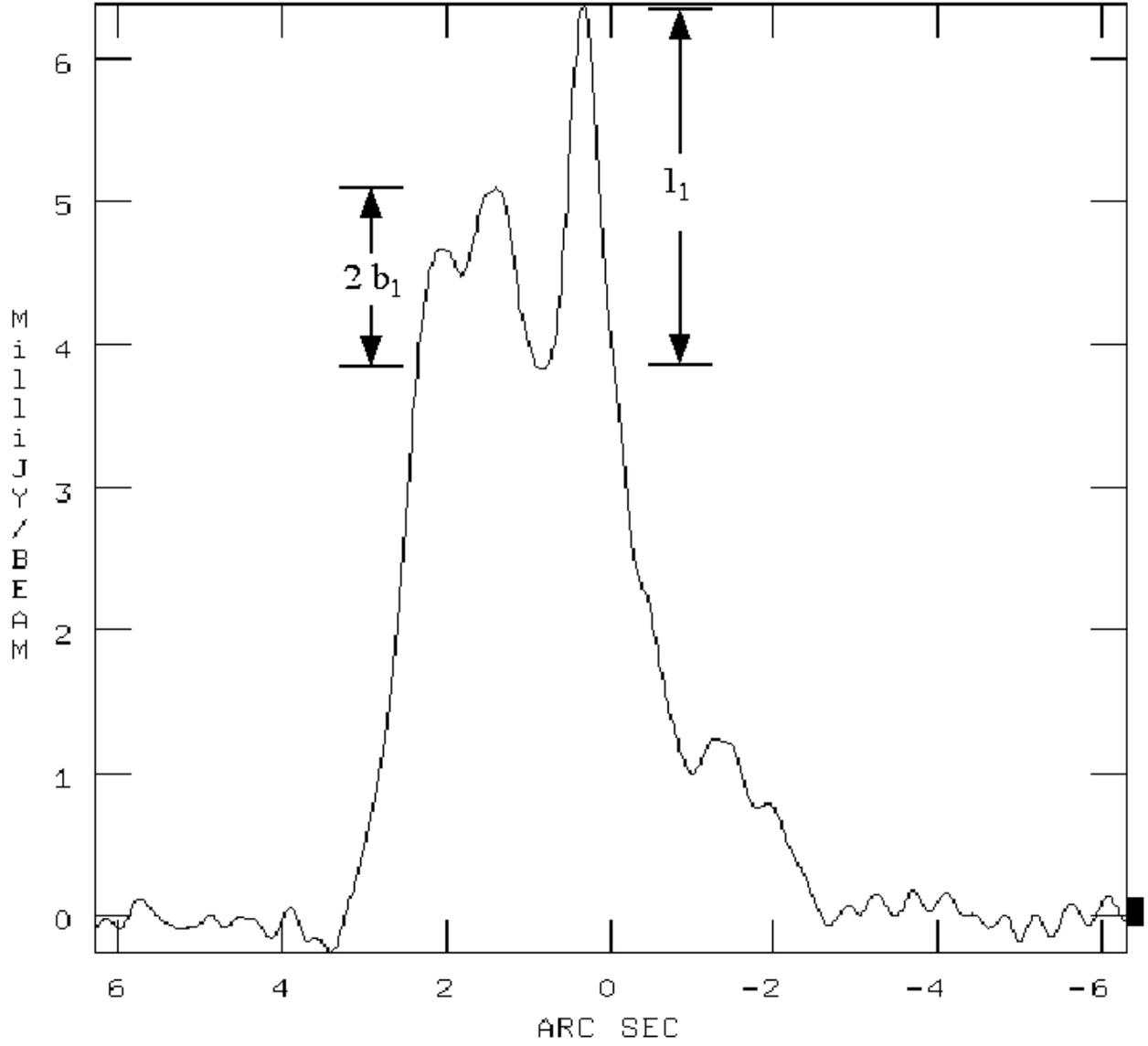}
}
\vspace{12pt}
\caption{Example of determining $I_1$ and $b_1$ for calculating the error in 
spectral index.  $I_1$ is conservatively estimated as 2.0 mJy, and $b_1$ is
estimated at 0.8 mJy.}
\label{modnoise}
\end{figure}

We thus took the above analytical errors as a reasonable estimate and
applied this to the real sources.  Using slices from the total
intensity maps at both frequencies, the intensities of the jet ($I_1$
and $I_2$) and irregularities in the lobe ($b_1$ and $b_2$) were
estimated by eye.  Since our models showed that irregularities with a
width up to twice that of the jet created the largest errors, any
irregularities on scales larger than this were ignored.  To be conservative, we
chose values for $b_1$ and $b_2$ equal to the largest deviations from
the smoother background lobe.  See Figure \ref{modnoise} for an
example of how these values were determined.  In this slice, the lobe
intensity is estimated to be $\approx 4.5$ mJy/beam.  This gives a
conservative estimate for the jet intensity of $\approx 2.0$ mJy/beam.  The
intensity of the ``noise'' that we modeled as a sine wave is estimated
as half of the total peak-to-peak deviation from the lobe - in this
case $\approx 0.8$ mJy/beam.

In one situation, the spectral index of the jet could not be determined
due to a third structure overlapping both the jet and lobe.  This
occurred in the northern lobe of 3C438 where a ring crosses the jet
about 5" from the core.

Rudnick (2001b) also determined the spectral index along the
jets of 3C438 and 3C401 using a multi-resolution filtering technique which
automatically removes the contributions from large scale structures.
The spectral index values found in that independent method agree quite
well with those found here using spectral tomography.

\subsection{Jet Properties} 

In order to compare the jet spectra with other jet properties, we
determined the FWHM and peak intensity of the jet, and the peak
intensity of the lobe by fitting a narrow and a wider gaussian to a
slice perpendicular to the jet at each location.  Occasionally, a
third narrow gaussian was used to improve the fit when a filament or
other structure was present in the slice.  The errors shown for the
width and intensity of the gaussians are formal errors in the fit. 
They do not take into account the fact that the jet and lobe are not
perfect gaussians and that intermediate size structures may also be
present.

\begin{figure}
\vskip -2in
\centerline{
\plotone{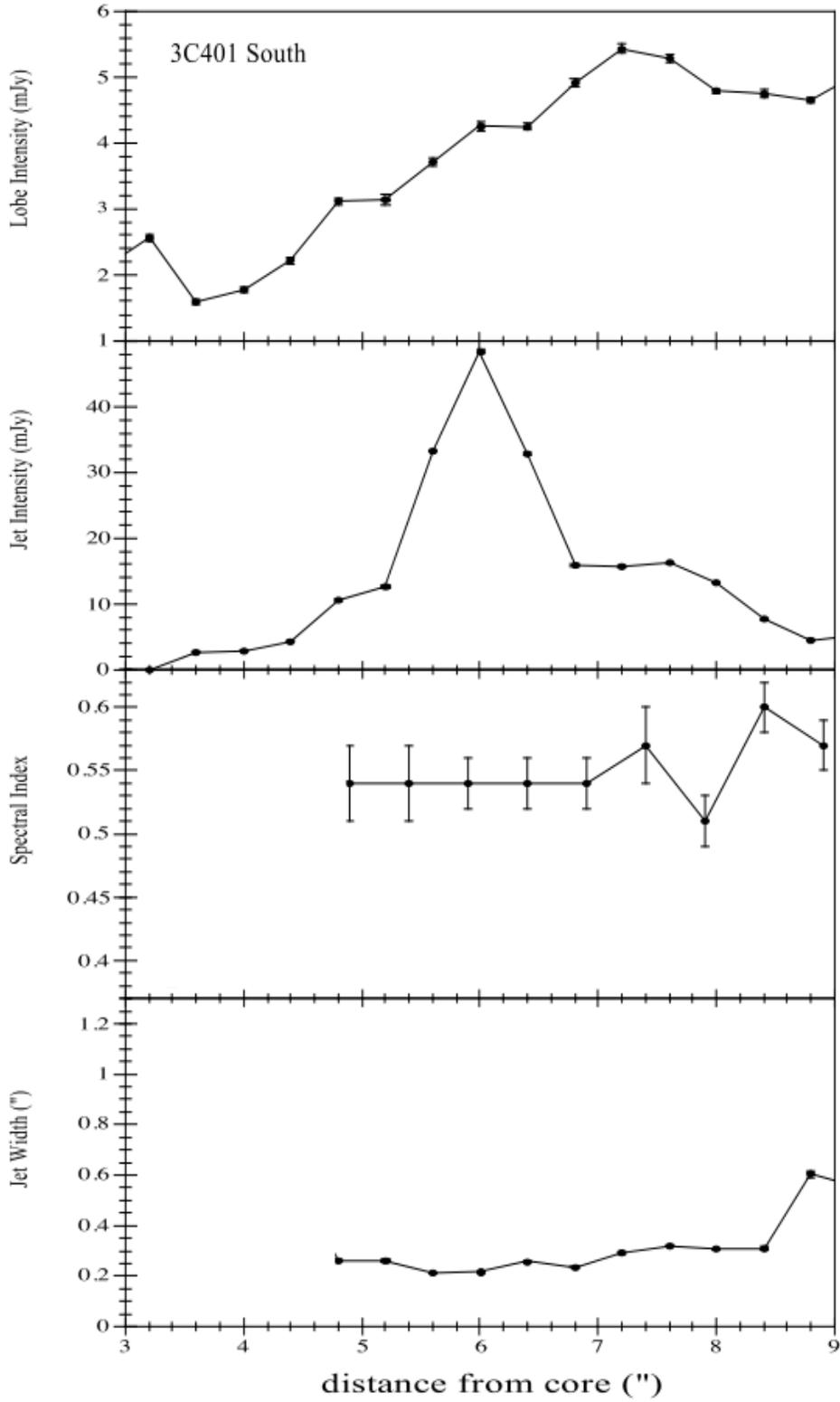}
}
\caption{3C401: distance vs: a) lobe intensity, b) jet intensity, c) spectral 
index of jet, and d) width of jet. The jet width is not corrected for the 0.25" beam 
at X band, so the first 7" are unresolved.} 
\label{slices401}
\end{figure}

\begin{figure} 
\centerline{
\plotone{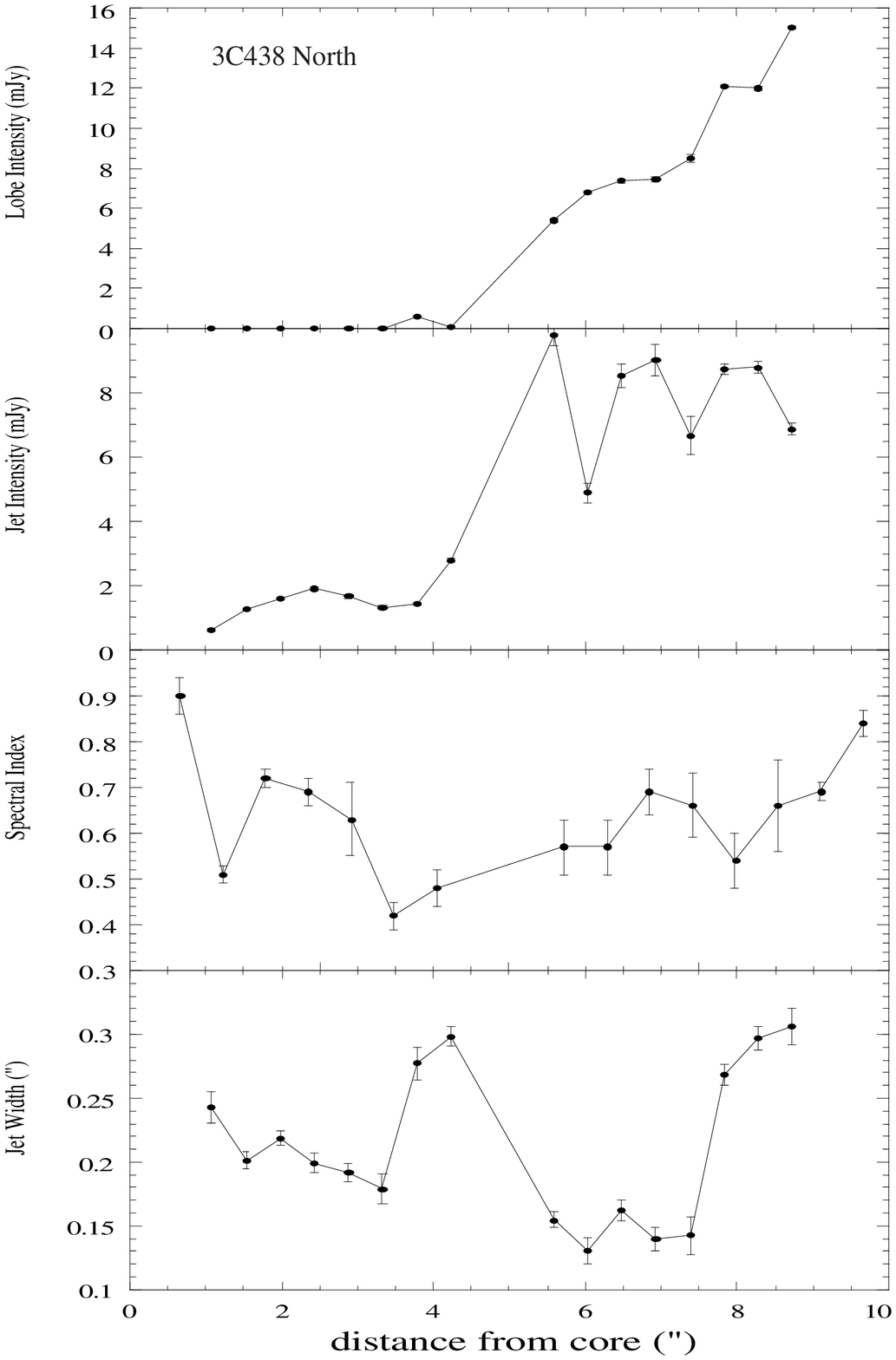}
}
\caption{Same as Figure \ref{slices401}, but for 3C438 north.}
\label{slices438N} 
\end{figure}

\begin{figure}
\centerline{
\plotone{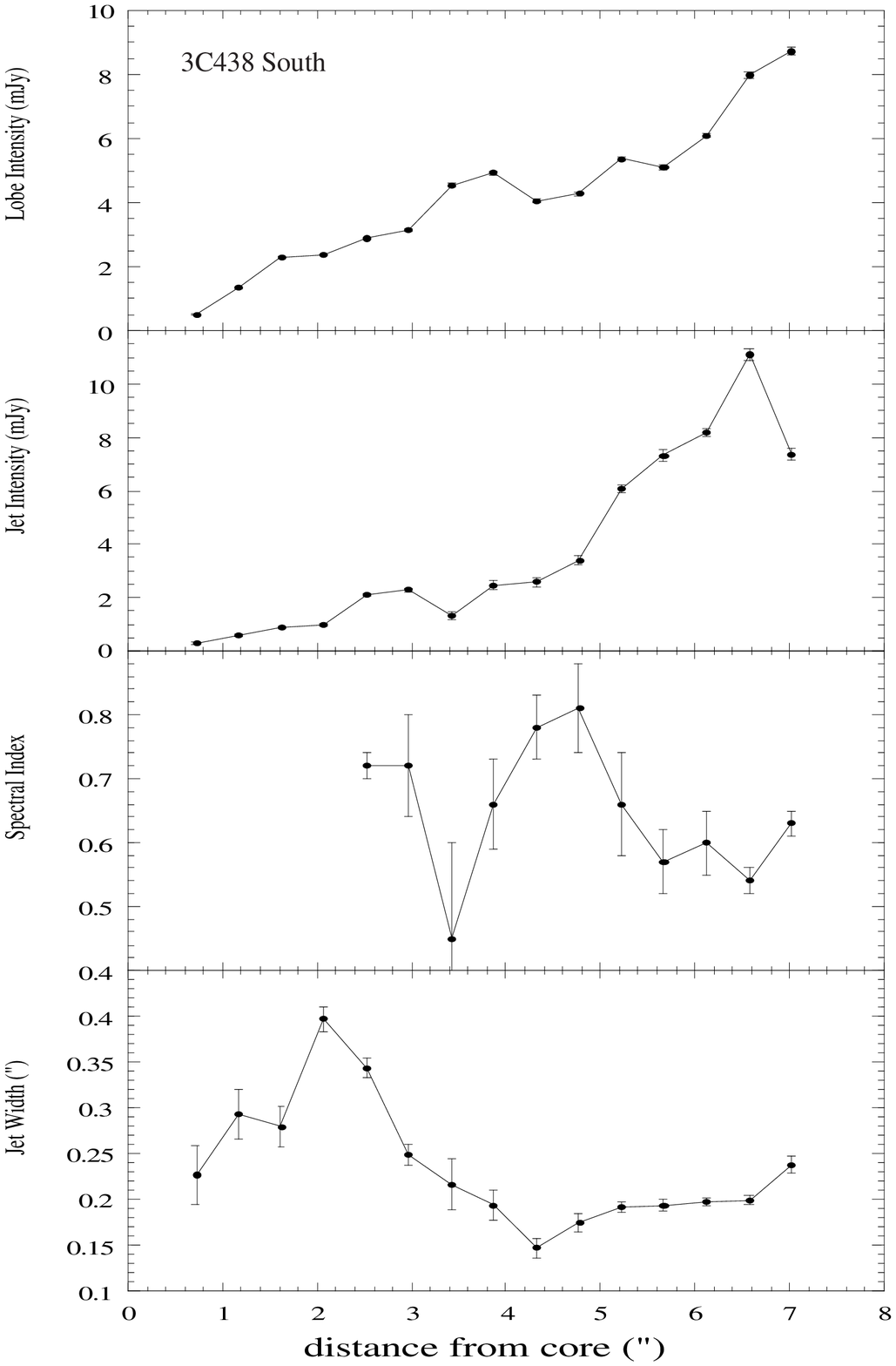}
}
\caption{Same as Figure \ref{slices401}, but for 3C438 south.}
\label{slices438S}
\end{figure}

Figure \ref{slices401} shows the results for 3C401.  In this jet, the
spectral index stays nearly constant at $\alpha=0.54$ along the length
of the jet. The jet width is unresolved by the Xband 0.27" beam
 until $\approx 7"$ from
the core.  Figures \ref{slices438N} and \ref{slices438S} are the graphs for the
northern and southern jets of 3C438, respectively.  In the northern
jet, the spectral index is flattest at approximately 3.5'' from the
core with almost monotonic steepening in both directions.  Just beyond
the point that the spectrum flattens, there is a dramatic widening of
the jet and an increase in the intensities of both the jet and the
lobe.  The steepest spectral indices in the southern jet are seen between
4''-5'' from the core.  This steepening happens at the same point that
the jet narrows in width - the opposite direction of the relation seen
in the northern lobe.

\subsection{Jet/Lobe Interactions}

When looking at the intensities of the jet and lobe in the northern
lobe of 3C438 (Figure \ref{slices438N}), we found that while the jet
can be seen well before the lobe emission, it then brightens considerably
as it enters the lobe.  Seeing this effect in one source prompted us
to look for similar examples in other sources.  We used the DRAGN
Atlas \footnote{Available at URL: http://www.jb.man.ac.uk/atlas/}
(Leahy et al.  1997) to identify a sample of FRII radio galaxies that
had a jet and a lobe that did not remain bright all the way back to the core,
i.e. there was a jet that ``entered'' the bright region of the lobe.  Of the fifteen jets
that fit this criteria, twelve had cleanly separable jet and lobe
emission enabling further study. 

\begin{figure}
\centerline{
\plotone{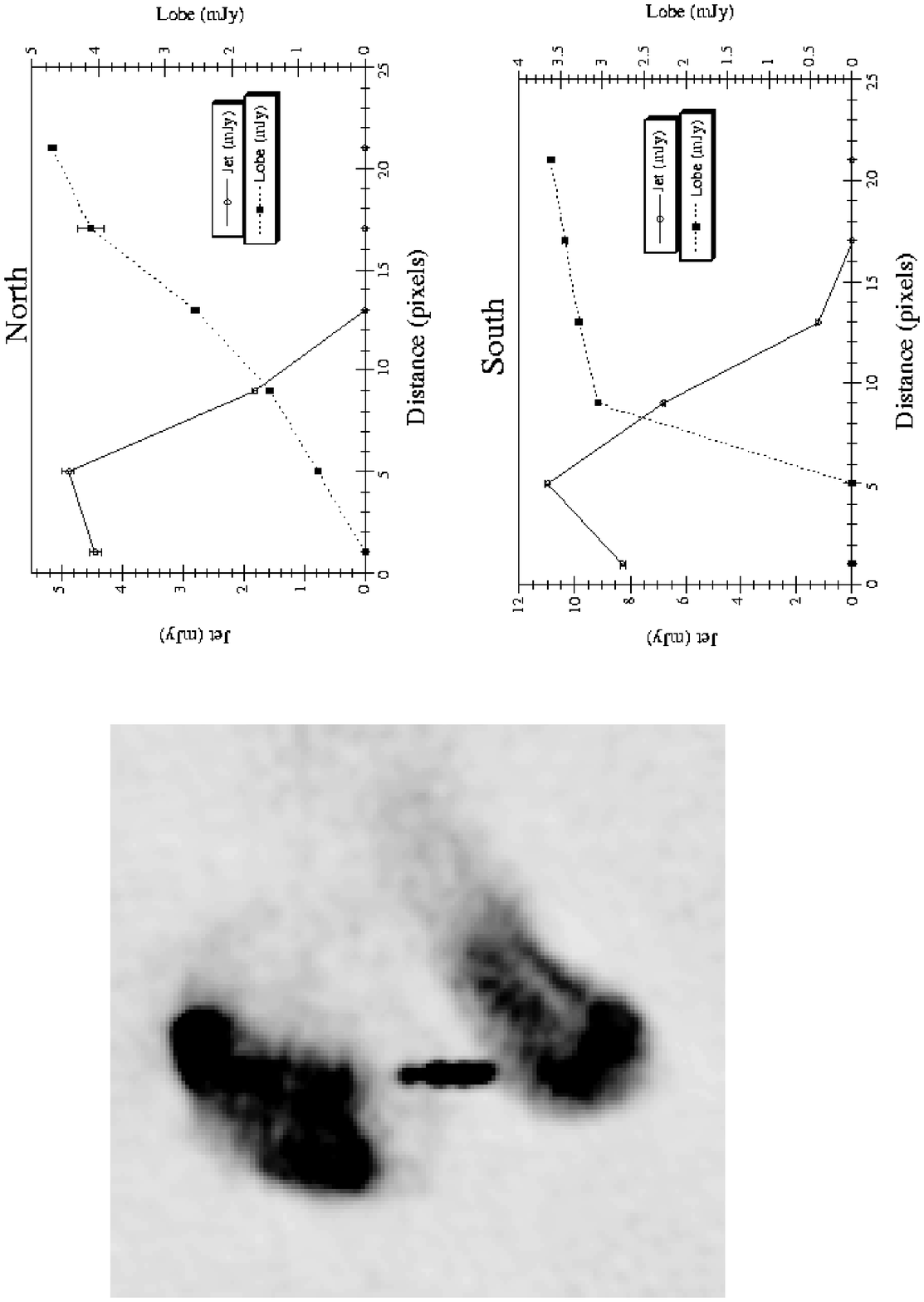}
}
\caption{a) Grey scale image of 3C288.  Intensity of jet and lobe for the
b) north and c) south lobe.}
\label{grey288}
\end{figure}

Our sample of twelve jets (see Table 2)
included nine which showed a significant change in jet brightness as
they entered the bright lobe and three (3C200, 3C305, and 3C401) which
showed no significant change.  In one source, 3C288, the jet faded
after it entered the lobe (see Figure \ref{grey288}), while in the rest
of the sources the jets brightened after entering the bright region
of the  lobe.

\section{Discussion}

\subsection{Mixed Spectral Indices}

The basic model of FRII radio galaxies includes lobes whose spectra gradually
steepen with distance from the hot spots due to aging of the emitting
electrons.  In the lobes of 3C438, 3C401, and 3C153, the most diffuse
parts of the emission follow this trend.  However, we also see a
complicated mixture of steep and flat spectral indices within the
lobes.  This mixture is seen in numerical simulations by Jones et
al. (1999).  These simulations show that electrons may encounter a
variety of shock types and strengths as they emerge from the jet,
creating different pockets of steep and flat spectrum electrons.  This
implies that these pockets of steep and flat spectra represent the
{\it history} of particle acceleration of these electrons and do not
necessarily reflect a {\it current site} of particle acceleration.

The ``c''-shaped, flat-spectrum structure seen in the southern lobe of
3C438 may also be a past termination point for a flapping jet.  Such a
jet has been suggested by Hardcastle et al. (1997) and is seen in numerical
simulations (e.g. Norman, 1996).

It is curious that our two best examples of mixed spectral indices,
3C438 and 3C401, are both in high density environments.  (They are
found in clusters, are surrounded by faint, diffuse x-ray emission, and
have unusually low polarization, indicating a dense magneto-ionic
environment, Hardcastle et al., 1997.)  But is it also possible that the dense
environment in some way caused the complicated spectral index
structures that we see?  Further comments about the interesting 
structures in these sources can be found in the DRAGN Atlas.

The gradient in spectral index that we see in several of our sources is
also seen in numerical simulations by
Ryu \& Jones (private communication).  In their
simulations, the spectral index of the hot spots varied from 0.5 to 1.0
due to a strong magnetic field found at the end of the jet which caused
severe spectral steepening.  The hot spots in our study had spectral
indices that ranged from 0.68 to 1.28, but each individual hot spot
only varied up to 0.2.

\subsection{Structures Parallel to Jet}

There is a growing number of structures seen parallel to jets, as we
found in 3C438.  Other than their proximity and structural
similarities to the jets, there is currently no way to separate these
structures from random filamentary features in the lobes.  Some
parallel structures are seen in polarization (e.g. 3C66B, Hardcastle
et al., 1996 and 3C219, Clarke et al., 1992) while others are seen in
total intensity maps (e.g. Cygnus A,
Katz-Stone \& Rudnick, 1994).  For 3C353 (Swain et al., 1998),
structures parallel to the jets are seen in both polarization and
total intensity.  These structures seen in FRIs may also be related to
the steep ``cocoons'' found around jets in WATs (Katz-Stone \&
Rudnick, 1994; and Katz-Stone et al., 1999).  It is not clear how
these various parallel structures are related to each other or to the
particular example found in 3C438.

One interpretation of these features is as a slower moving sheath
surrounding a faster spine, or inner jet (Laing, 1996).  Spine/sheath
models of jets have been shown to stabilize jets and may dominate jet
dynamics (Sol et al., 1989 and Hanasz \& Sol, 1996).  If the sheaths
are slower moving, then at a given distance from the core the
electrons in the sheath could have aged and steepened more if the
magnetic fields in the jet and sheath are comparable.  Alternatively,
the same curved electron energy distribution could appear steeper in
the sheath due to lower magnetic fields (Katz-Stone et al., 1999).
While it may be possible that the parallel structure we see in 3C438
is unrelated to the jet, this appears unlikely due to the close
correspondence in their structure and brightness.

\subsection{Spectra of Jets}

The spectra of jets should reflect both the relativistic particle
acceleration and loss histories.   In WATs and other lower
luminosity sources, the radio spectral index appears to be constant along
the inner jet (3C449, Katz-Stone \& Rudnick, 1997b: $\alpha= 0.53 $;
 3C66B, Hardcastle et al., 1996: $\alpha= 0.5~  to~  0.6 $; 
M87, Owen et al., 1989: $\alpha= 0.5~  to~  0.6 $.)
This suggests either that radiative losses are unimportant in the
radio, or that the acceleration process is important all along the
jet, and robust to the physical variations causing brightness and width changes.
Even the radio/optical spectral indices for 3C66B (Jackson et al. 1993) or the
derived power laws for M87 (Meisenheimer et al., 1996) show little variation
along the jets. The strongest support for acceleration distributed along
the jet comes from their  X-ray detection, (e.g. 3C66B, Hardcastle, Birkinshaw
and Worrall 2001), for which the lifetime of the presumed
 synchrotron emitting electrons can be as short as 30 years.

Very little information exists on the spectra of FRII jets. 3C236 shows considerable
variation in the  radio spectral index ($\alpha= 0.5~  to~  1.5$)  of its inner components
(Schillizzi et al. 2001).  Cygnus A's jet shows a similarly steep spectrum and significant curvature
up to $\lambda 2cm$, with possible variation along the jet that is difficult to
separate from the background (Katz-Stone, Rudnick \& Anderson 1993).
As noted earlier, we find that 3C401's jet has an approximately constant
spectral index of $\alpha = 0.53$, similar to the FRI results above. 
We do see spectral variation in both jets of 3C438, but
these are not monotonic steepenings as one might expect from simple
radiative losses.

The origin of the spectral variations in the FRII jets is not clear.
We looked for clues from the variations of spectrum, width and 
intensity in 3C438's jets.
In the southern jet, we found that the spectral index steepens at the
same point that the jet narrows.  In this same sense, in the northern
jet the spectrum flattens as the jet widens.  There are several
possible explanations for this phenomenon among which we are unable to
distinguish at this time. One possibility is that there is no evolution
of the particle spectrum along the jet, but that varying magnetic fields
result in different local spectral indices.  An arbitrary tuning of the
relativistic electron densities  would then be required to fit the local
brightnesses. Another possibility is that particle acceleration is
occurring along the jet, and varies with the local conditions.  For example,
 the increase in jet
width could be accompanied by increased jet turbulence which could also
increase particle acceleration,  flatten the spectrum and increase the intensity.   A
dramatic increase in intensity is seen immediately after the spectrum
flattens at $\approx$3.5'' from the core in the northern lobe of
3C438.  If there is particle acceleration at this point,
multi-frequency observations should show a power law spectrum with a
high break frequency.

We  cannot  draw conclusions from a single example, but the literature on FRII jets  provides
little additional guidance.  Cygnus A's jet shows regions of expansion and
constant width (Perley, Dreher \& Cowan 1984), although the connection to
its knotty brightness profile is unclear.   3C219 shows significant variations in jet width, which
Bridle, Perley \& Henriksen (1986) suggest may be related to its local
brightening. 3C353 has a well resolved jet width that remains virtually constant,
although it has significant changes in brightness (Swain, Bridle \& Baum 1998).
Jet widths of four quasars are shown by Bridle et al. 1994, and have similar
patterns of widening and narrowing as seen here for 3C438.
 These various behaviors, and the different possible interpretations of
3C438's behavior, reflect very different
physical pictures for the jet  and the supply of relativistic particles
to the hot spots and lobes. Sorting this out is likely to depend on 
multifrequency observations to measure local spectral shapes, with sufficient
resolution to separate knots from the  underlying structure in the jets.

\subsection{Jet/Lobe Interactions}

As discussed above, in nine of the twelve cases where we could isolate
the jet and lobe intensities, we found significant changes in these
jets as they entered the bright part of the lobe.  How does the lobe
affect the jet?  One possibility is that the turbulence in the lobe
may disrupt the jet and cause it to lose its coherency; the jet will
decrease in brightness as seen in 3C288. Or the lobe material may
stimulate shocks in the jet (Jones, Ryu \& Engel, 1999), increasing its
brightness as we see for the majority of the sources we studied.
 Another possibility is that
when the jet enters the lobe, a second-order Fermi process
reaccelerates  particles in the boundary layer between the jet and
lobe.  If the increased  brightness is caused by shocks
stimulated by lobe material or by a second-order Fermi process in the
boundary layer, the jet's spectrum should flatten as it enters the
lobe.  This flattening is a signature of particle acceleration.  For
the one jet in our study which enters the lobe and which we have
spectral index information (3C438 North), we see a slight, but
inconclusive, flattening of spectral index.

\section{Concluding Remarks}

The relevance of the simple cartoons of powerful radio galaxies
has eroded over the years.  The jets are not steady 
(e.g., Burns, Christiansen \& Hough 1982; Clarke \& Burns 1991);
indeed they may flap around near their terminus (Scheuer 1982).
as indicated by both observations (Hardcastle et al. 1997)
 and numerical simulations (Hardee \& Norman 1990). 
As a consequence, the canonical steady terminal shock (Blandford
\& Rees 1974), at 
which relativistic particles can be accelerated, does not exist.
This raises the question about where particle energy gains and
losses actually occur in radio galaxies.  This question is
highlighted by  recent optical and X-ray observations of jets (M87,
Perlman et al. 1999; Pictor A, Wilson, Young \& Shepbold 2001)
which may require  particle acceleration  along the entire 
length of the jet. The nature of the broad but ``jet-like'' structure
in Cygnus A's X-rays is unknown (Wilson, Young \& Shepbold 2001).
 As far as simple pictures of the backflowing
cocoons, we instead see that  backflows are not smooth, and lobes may be 
populated by filamentary features (Owen, Eilek and Kassim 2000;
Swain,  Bridle \& Baum 1998;  Katz-Stone, Rudnick, \& Anderson 1994).
These  may indicate flow 
instabilities, shocks or turbulence from interaction with the jet and 
surrounding medium.  In the weaker, FRI, sources, there are clear 
indications of stratified jets, or jet/sheath structures (Katz-Stone et al. 1999). 

This new level of complexity is well-illustrated in the latest 
generation of numerical simulations (Tregillis, Jones \& Ryu 2001)
which include relativistic particle acceleration along with 3D MHD.
At issue is not simply the ``weather'' of radio galaxy structure, but
the fundamental questions of what parts of the outgoing flow are 
actually visible as the ``jet'', how the unstable jet drives the 
overall source dynamics and structure, and where and how the bulk of 
the relativistic particle acceleration takes place.  

The spectral tomography work presented here illustrates the type of
information available and needed to address such issues.  Further work 
using high-quality data is especially needed to follow up on a) the 
spectral index structure in hot spot regions, b) spectral index 
variations along jets, and c) possible jet/lobe interactions. At the 
same time, broad explorations of parameter space with 
numerical simulations are needed to determine under what conditions
behaviors such as reported here are found.

\acknowledgements 

We would like to thank Robert Laing and Rick Perley for
the use of their data.  MJH acknowledges support from PPARC grant
number GR/K98582.  This work is partially supported at the University of
Minnesota by grants AST96-16964 and AST00-71167
 from the National Science Foundation.

\clearpage

\clearpage

\clearpage

\clearpage

\begin{table}
\caption{Frequency, resolution, and noise levels for the sources
studied.  References - (1) Laing, private communication; (2)
Hardcastle et al., 1997; (3) Leahy et al., 1997 (Available at URL:
http://www.jb.man.ac.uk/atlas/; (4) Leahy \& Perley, 1991. }
\begin{center}
\begin{tabular}{|r|r|r|r|l|}\hline
Source 	& Frequency 	& Beamsize 	& RMS Noise  & Reference \\ 
	& (GHz)  	& ('') 		& ($\mu$Jy/beam) & \\ \hline
3C132 	& 4.96 	& 0.33 	& 38 	& 1 \\
 	& 8.44 	& 0.22 	& 29 	& 2 \\ 	\hline

3C153 	& 1.53 	& 0.15 	& 95	& 3 \\
	& 8.44	& 0.25	& 28	& 2 \\ 	\hline

3C173.1	& 1.45	& 2.60	& 77	& 4 \\
	& 8.44	& 2.60	& 89	& 2 \\		\hline

3C349	& 1.50	& 2.90	& 102	& 4 \\
	& 8.44	& 2.90	& 59	& 2 \\		\hline

3C381	& 1.45	& 3.25	& 151	& 4 \\
	& 8.44	& 3.25	& 99	& 2 \\		\hline

3C401	& 1.53	& 0.35	& 76	& 3 \\
	& 8.44	& 0.27	& 24	& 2 \\		\hline

3C438	& 1.53	& 0.29	& 97	& 3 \\
	& 8.44	& 0.23	& 25	& 2 \\		\hline

4C14.11	& 1.45	& 3.00	& 68	& 4 \\
	& 8.44	& 3.00	& 29	& 2 \\		
\hline
\end{tabular}
\end{center}
\end{table}

\clearpage

\begin{table}
\caption{Jet change in brightness upon entering
lobe.  Notes: (1) The linear feature in the lobe, which is a straight
extension from the jet, could be a filamentary feature such as
seen in 3C338, (Ge \& Owen 1994)  whose nature is still unclear.
(2) Although lobe shows also shows a transition where the jet brightens,
faint lobe material extends all the way back to the nucleus and
becomes confused with lobe material from the opposite side.
}
\begin{center}
\begin{tabular}{|l|c|c|l|}\hline

Source	& Number of	& Change in Jet	& Reference \\ 
	& Jets		& Brightness?	&	\\	\hline
3C28	&1	&Increase(1) 	&Leahy, Bridle \& Strom, 1997		\\
3C33.1	&1	&Increase(2) 	&Leahy, Bridle \& Strom, 1997		\\
3C200	&1	&No		&Clarke, D. A. and Burns, J. O. (unpublished)\\
3C219 	&1	&Increase(2)	&Clarke et al., 1992	\\
3C285	&1	&Increase	&Leahy et al., 1986	\\
3C288	&2	&Decrease	&Bridle et al., 1989	\\
3C305	&1	&No		&Leahy, Bridle \& Strom, 1997 \\
3C388	&1	&Increase(2)	&Roettiger et al., 1994	\\
3C401	&1	&No		&Hardcastle et al., 1997; Leahy, Bridle \& Strom, 1997\\
3C438	&2	&Increase	&Hardcastle et al., 1997; Leahy, Bridle \& Strom, 1997\\
\hline
\end{tabular}
\end{center}
\end{table}
\clearpage


\clearpage


\begin{thebibliography}{}
\centerline {}
\bibitem {bl84}Blandford, R.D. \& Rees, M.J., 1984, Reviews of Modern Physics, 56 255

\bibitem{bo01}Bohringer, H. et al. 2001, \aap, 365 181

\bibitem {b89}Bridle, Alan H. et al. 1989, \aj  97 674 

\bibitem {b86} Bridle, A. H., Perley, R. A. \& Henriksen, R. N. 1986 \aj  92 534

\bibitem{bu82} Burns, J. O., Christiansen, W. A. \& Hough, D. H. 1982 \apj, 257 538

\bibitem{c01}Celotti, A., Ghisellini, G. \& Chiaberge, M., 2001 \mnras, 321 L1

\bibitem{c92}Clarke, David A. et al. 1992, \apj, 385 173 

\bibitem{c91}Clarke, D. A. \& Burns, J. O. 1991, \apj, 369 308

\bibitem{c83}Cornwell, T., 1983, \aap  121 281

\bibitem{e91} Eilek, J. A. and Hughes, P. A.  1991 in Beams and Jets in
Astrophysics P.  Hughes, Cambridge: Cambridge University Press 428



\bibitem{ge94} Ge, J. \& Owen, F.N. 1994, \aj  108, 1523

\bibitem {f84}Feretti, L. et al. 1984, \aap  139 50 

\bibitem{h96} Hanasz, M. and Sol, H. 1996, \aap,  315  355 

\bibitem{mjh96} Hardcastle, M. J. et al. 1996, \mnras,  278  273 

\bibitem{mjh97} Hardcastle, M. J. et al. 1997, \mnras,  288  859 

\bibitem{mjh01} Hardcastle, M. J. Birkinshaw, M. \& Worrall, D. 2001, (preprint) 

\bibitem{hxx} Hardee, P. \& Norman, M. L. 1990, \apj  365 134

\bibitem{j93} Jackson, N., Sparks, W. B., Miley, G. K., \& Macchetto, F. 1993, \aap, 269 128

\bibitem{j2k} Jones, T. W., Ryu, D. \& Engel, A., 1999, \apj  512 105

\bibitem{k94} Katz-Stone, D. M., and Rudnick, L. 1994, \apj,  426  116 

\bibitem{k97a} Katz-Stone, D. M., and Rudnick, L. 1997a, \apj,  479  258 

\bibitem{k97b} Katz-Stone, D. M., and Rudnick, L. 1997b, \apj,  488  146 

\bibitem{k99} Katz-Stone, D. M., Rudnick, L., Butenhoff, C. \& O'Donaghue, A., 1999,
\apj, 516 716

\bibitem{l96} Laing, R.A. 1996 in Energy Transport in Radio Galaxies and
Quasars ASP Conference Series Vol. 100 P. E.  Hardee, A. H. Bridle and
J. A.  Zensus, Provo: Brigham Young University Press, 241

\bibitem{l91} Leahy, J. P. 1991 in Beams and Jets in Astrophysics P.
Hughes, Cambridge: Cambridge University Press 100

\bibitem {l86}Leahy, J. P. et al. 1986, \mnras   222 753 

\bibitem{l97} Leahy, J. P., Bridle, A. H. \& Strom, R. G. 1997, ``An Atlas of DRAGNs,"
  http://www.jb.man.ac.uk/atlas

\bibitem{lp91} Leahy, J. P., and Perley, R. A. 1991, \aj,  102  537 

\bibitem{96m} Meisenheimer, K. et al. 1996, \aap,  307  61 

\bibitem{n96} Norman, M. L. 1996 in Energy Transport in Radio Galaxies and
Quasars ASP Conference Series Vol. 100 P. E.  Hardee, A. H. Bridle and
J. A.  Zensus, Provo: Brigham Young University Press, 319

\bibitem{o89} Owen, Frazer N. et al. 1989, \apj,  340  698 

\bibitem{o00} Owen, F.N., Eilek, J. A. \& Kassim, N. E., 2000, \apj, 543 611

\bibitem {p84} Perley, R. A., Dreher, J. W. \& Cowan, J. J. 1984 \apj  285 L35

\bibitem {r94}Roettiger, Kurt et al. 1994, \apj  421  L23 

\bibitem{r01a} Rudnick, L. 2001, ``What Shape are Your Spectra In?" in Proceedings of the Oxford Radio Galaxy Workshop,
eds. R. Laing \& K. Blundell,  ASP Conf. Series (in press)

\bibitem{r01} Rudnick, L. 2001b, ``Simple Multiresolution Imaging and the Spectra of
 Synchrotron Sources", submitted to \apj 

\bibitem{s82}Scheuer, P.A.G. 1982, in Extragalactic Radio Sources, Proc. of IAU Symp. 97,
Dordrecht, D. Reidel, 163

\bibitem{sch01}  Schilizzi, R. T.,  Tian, W. W.,  Conway, J. E.,  Nan, R.,  Miley,
G. K.,  Barthel, P.D.  Normandeau, M., Dallacasa, D. \& Gurvits, L. I., 2001 \aap, 368 398

\bibitem{s89}Sol, H. et al. 1989, \mnras,  237  411 

\bibitem{s98} Swain, Mark R., Bridle, A.H., \& Baum, S. 1998,  \apj,  507  29 

\bibitem{t01} Tregillis, I., Jones, T.W. \& Ryu, D., 2001, \apj  (in press, astro-ph/0104305)

\bibitem{w01} Wilson, A., Young, A. J., \& Shopbell, P.L., 2001, preprint (astro-ph/0101422)

\end{thebibliography}
\end{document}